\documentclass[aps,prd,10pt,twocolumn]{revtex4}

\usepackage{amssymb}
\usepackage{amsmath}
\usepackage{amsfonts}
\usepackage{graphicx, float}
\usepackage{graphicx, epsfig}
\usepackage{color}
\usepackage{enumerate}

\begin{document}

\title{Thermodynamics, shadow and quasinormal modes of black holes\\ in five-dimensional Yang-Mills massive gravity}

\author{S. H. Hendi$^{1,2}$\footnote{email address: hendi@shirazu.ac.ir} and
A. Nemati$^{1}$\footnote{email address: j.azadeh.nemati@gmail.com}
} \affiliation{$^1$Physics Department and Biruni Observatory,
College of Sciences, Shiraz
University, Shiraz 71454, Iran\\
$^2$Canadian Quantum Research Center 204-3002 32 Ave Vernon, BC
V1T 2L7 Canada}

\begin{abstract}
In this paper, we consider the Einstein-massive gravity coupled to
the Yang-Mills gauge field in five dimensions. Concentrating on
the static solutions with planar horizon geometry, we explore the
thermodynamic behavior in the extended phase space and examine the
validity of the first law of thermodynamics besides local thermal
stability by the Hessian matrix. We observe that although the
topology of boundary is planar, there exists the van der Waals like
phase transition for this kind of solution. We find the critical
quantities and discuss how the massive and Yang-Mills parameters
affect them. Furthermore, some signatures of the first order phase
transition such as the swallow-tail behavior of the Gibbs free
energy and divergencies of the specific heat are given in detail.
We continue with the calculation of the photon sphere and the
shadow observed by a distant observer. Finally, we use the WKB
method to investigate the quasinormal modes for scalar
perturbation under changing the massive and Yang-Mills parameters.
\end{abstract}

\maketitle

\section{Introduction}

One of the interesting predictions of general relativity is the
existence of black holes which are the most mysterious objects in
the Universe. Although it is believed that black holes are
invisible, it is possible to have an image of them; which has been
captured by the Event Horizon Telescope \cite{Akiyama:2019cqa}.
The image of a supermassive black hole in the galaxy $M87$, a dark
part which is surrounded by a bright ring, was the first direct
evidence of the existence of black holes consistent with the
prediction of general relativity. What makes it possible is
bending the light of luminous sources behind the black hole or
emitting radiation from matter plunging into the black hole. In
spite of the fact that gravitational lensing (bending of light
rays by spacetime curvature) is crucial, the existence of unstable
photon orbits makes it possible that photons in the vicinity of a
black hole escape from gravity to a distant observer, where they
form a bright ring around the black shadow (for a recent review see
\cite{Dokuchaev:2019jqq}).

Black hole's shadow, as an important observable, has become the
subject of interest, thereby we can extract information about the
parameters of a black hole and its near geometry
\cite{Bambi:2019tjh}. Moreover, the shadow image measurements can
determine whether Einstein's general relativity is correct or
whether it should be modified in the presence of strong
gravitational fields \cite{Moffat:2015kva}. One of the goals of
this paper is to explore the shadow of the $5-$dimensional
massive-Yang-Mills black hole and study the effect of different
parameters on the size of photon orbits and spherical shadow.

Black holes are interacting with the matters and radiations in
surroundings. The gravitational waves are the signals of black
holes in reaction to these perturbations \cite{GW}. The dominant
part of such signals are quasinormal modes which are damping
oscillations after the initial outbursts of radiation and they end
to the late-time tails after a very long time. The gravitational
quasinormal modes can determine if the black hole is stable and
thus may exist in nature. On the other hand, quasinormal modes are
connecting with the thermodynamic phase transition in the dual
field theory according to the AdS/CFT correspondence and the
hydrodynamic regime of strongly coupled field theories (for a
detailed discussion see \cite{Konoplya:2011qq}). So we are
motivated to investigate quasinormal modes for the considered
black hole. For some works calculating quasinormal modes for
solutions of dRGT massive gravity in four dimensions see
\cite{Ponglertsakul:2018smo,Hendi:2018hdo,Moulin:2019ekf,Zou:2017juz,Burikham:2017gdm,Chen:2019kaq,Prasia:2016fcc}.

In the path of finding new physics, black holes play an essential role. Among the huge range of studies about the
different aspects of these objects, observing similarity with the
everyday thermodynamics is certainly astonishing
\cite{Kubiznak:2016qmn}. As an interesting similarity, one may
find a van der Waals like behavior for spherically symmetric
charged AdS black holes \cite{Kubiznak:2016qmn}. This becomes
possible with changing the role of black hole mass and defining a
thermodynamic pressure term proportional to the cosmological
constant. Correspondence between the black hole parameters and the
fluid thermodynamic quantities, in this perspective, gives rise to
a new picture of black holes called black hole chemistry. While
the small-large black hole phase transition now mimics the role of
liquid-gas phase transition. Regarding the thermodynamic pressure
with thermodynamic volume as its conjugate quantity guides us
towards the generalized first law in the extended phase space
thermodynamics. Investigation of thermodynamic phase transition in
the context of alternative theories of gravity is interesting,
since one may encounter a new phenomenon which is not observed in
Einstein gravity (for example see \cite{Hendi:2017fxp}).

One of the highly successful extensions of the Einstein gravity is
dRGT massive gravity \cite{deRham:2010ik}. This theory introduces
a novel potential term by the contribution of some polynomial
interactions and a reference metric which may lead to the mass
term for the graviton. This version of massive gravity gives
unique features to the theory and avoids the common instabilities
observed in other theories with a massive spin-$2$ field in their
spectrum \cite{deRham:2010kj,deRham:2014zqa}. The very basic
question that how the existence of mass for the graviton affects
the thermodynamic behavior of black holes has been explored in the
literature (see for e.g.
\cite{Hendi:2017fxp,Cai:2014znn,Hendi:2015eca,Ghosh:2015cva,Hendi:2016vux,Zou:2016sab}
and references therein). In this paper, we also consider black
hole solutions of the dRGT massive gravity coupled with the
Yang-Mills field and investigate their thermodynamics.

One of the interesting ways for coupling matter to the gravity theory is
using the non-abelian Yang-Mills gauge field. Criticality and possibility of
phase transition have been discussed for black holes in the theories
AdS-Maxwell-power-Yang-Mills \cite{Zhang:2014eap, ElMoumni:2018fml},
Einstein-Born-Infeld-Yang-Mills \cite{Meng:2017srs}, $f(R)$ gravity coupled
with Yang-Mills field \cite{Ovgun:2017bgx}, Gauss-Bonnet massive gravity
coupled to Maxwell and Yang-Mills fields \cite{Meng:2016its} and Wu-Yang
model of Yang-Mills massive gravity in the presence of Born-Infeld nonlinear
electrodynamics \cite{Hendi:2018hdo}.

Black hole thermodynamics may have different behavior in higher
dimensions. As an example, it was shown that \cite{Hendi-Dehghani}
for charged topological black holes in the grand canonical
ensemble, the van der Waals like phase transition could take place
only for $dimensions \geq 5$. The five-dimensional spacetime is
important for different aspects; from its verification as a
candidate for unifying electromagnetic and gravity forces to
appearing black hole solutions with different horizon topologies
\cite{Horowitz:2012nnc, Wesson:2014raa}.

In this paper, we focus on the Einstein-Massive gravity in the
presence of the Yang-Mills gauge field in $5$-dimensions.
Recently, the static solutions with the planar horizon for this
theory have been introduced in Ref. \cite{Sadeghi:2018ylh}. We are
interested to discuss the thermodynamic aspects of these solutions
in the extended phase space thermodynamics and looking for the
possible van der Waals like phase transition. Furthermore, we
investigate the possibilities for the formation of shadow and
variation of its size under considering different parameters. At
last, we calculate frequencies of quasinormal modes for scalar
perturbations by using the WKB method.

The plan of the paper is as follows: In Sec. \ref{sec2}, we give a
brief review of the action with the planar AdS solutions. Then, we
obtain the thermodynamic quantities. Investigation of the critical
behavior and the van der Waals phase transition, as well as
studying the effects of both massive and Yang-Mills parameters on
the critical quantities are addressed in Sec. \ref{sec3}. Next,
the consistency of local thermal stability conditions in two
canonical and grand canonical ensembles is discussed. Moreover, it
is shown that the first law is valid for the kind of considered
solutions. We also explore the photon orbits near the black hole
and formation of shadow in Sec. \ref{sec 4}. In the last section
\ref{sec5}, the quasinormal modes for scalar perturbation are
presented. Finally, we give the concluding remarks in Sec.
\ref{secCon}.
\section{Action and thermodynamics}

\label{sec2}

The action of $5-$dimensional Einstein-massive gravity with
negative cosmological constant in the presence of Yang-Mills
source is as follows
\begin{equation}  \label{action}
S=\int d^{5}x \sqrt{-g} \left(R-2 \Lambda - F^{(a)}_{\mu\nu}F^{(a) \mu\nu}
+m^2 \sum_{i=1}^{3} c_{i} \mathcal{U}_{i}(g,f)\right),
\end{equation}
where $R $ and $\Lambda $ are, respectively, the scalar curvature and the
negative cosmological constant. The Yang-Mills tensor $F^{(a)}_{\mu\nu} $ is
\begin{equation}
F^{(a)}_{\mu\nu}= \partial_{\mu} A_{\nu} - \partial_{\nu} A_{\mu}+ \frac{1}{%
2 e} f_{(b)(c)}^{(a)} A_{\mu}^{(b)} A_{\nu}^{(c)},
\end{equation}
where $e $ is the gauge coupling constant and $A_{\mu}^{(a)} $ is the gauge
potential. The $f_{(b)(c)}^{(a)} $ are gauge group structure constants and
for the $SU(N) $ group, one has $a, b, c= 1...N^2-1 $. Regarding the last
term of the action (massive term), $c_i$'s are some constants and $\mathcal{U%
}_{i}$'s are symmetric polynomials of the eigenvalues of $\mathcal{K}%
^{\mu}_{\nu}=\sqrt{g^{\mu\alpha} f_{\alpha\nu}} $ with the following
explicit forms
\begin{eqnarray}
&&\mathcal{U}_{1}= [\mathcal{K}], \\
&&\mathcal{U}_{2}= [\mathcal{K}]^2 - [\mathcal{K}^2],  \nonumber \\
&&\mathcal{U}_{3}= [\mathcal{K}]^3 -3 [\mathcal{K}] [\mathcal{K}^2]+ 2[%
\mathcal{K}^3].  \nonumber
\end{eqnarray}

We follow the reference metric ansatz $f_{\mu\nu}=(0,0,c_0^2
h_{ij}) $ with positive constant $c_0 $ and
$h_{ij}=\frac{1}{b^2}\delta_{ij} $, where ``$b$" is an arbitrary
constant with dimension of length. Hereafter, we follow the metric
solution of the five-dimensional planar AdS black brane introduced
in Ref. \cite{Sadeghi:2018ylh}
\begin{equation}  \label{metric}
ds^2= -f(r) dt^2 +\frac{dr^2}{ f(r)} +r^2 h_{ij} dx^{i} dx^{j},
\end{equation}
where by introducing notation $\mathtt{C}_i=m^2 c_0^i c_i$ with $i=1,2,3 $,
the blackening factor with the horizon radius $r_+ $ and $AdS $ radius $\ell=%
\sqrt{-6/\Lambda} $ is
\begin{eqnarray}  \label{black}
f(r)&=&\frac{1}{ r^2} \left[ \frac{r^4-r_+^4}{\ell^2} +\frac{\mathtt{C}_1}{3}
(r^3-r_+^3) + \mathtt{C}_2 (r^2-r_+^2) \right.  \nonumber \\
&&+\left. 2 \mathtt{C}_3 (r-r_+) -2 e^2 \ln (\frac{r}{r_+})
\right].
\end{eqnarray}

As we expect, in the limit of vanishing parameters for massive and
Yang-Mills terms, the familiar form of metric function $f(r)= \frac{r^2}{%
\ell^2}-\frac{r_{+}^4}{r^2 \ell^2} $ for the $5$-dimensional planar AdS
spacetime is recovered.

Considering Eq. (\ref{metric}), we can compute the Kretschmann scalar
\begin{equation}  \label{RR}
R_{\alpha \beta \gamma \delta}R^{\alpha \beta \gamma \delta}=\left(\frac{%
d^{2}f(r)}{dr^2}\right)^2+6 \left(\dfrac{1}{r}\frac{d f(r)}{dr}\right)^2+12
\left(\dfrac{f(r)}{r^2}\right)^2.
\end{equation}
By applying  Eq. (\ref{black}), we find that there is a curvature
singularity at $r=0$ which is covered with an event horizon
$r_{+}$. So the solution can be interpreted as planar black brane.
The first term of Eq. (\ref{black}) is dominant for large ``$r$"
reflecting the fact that the asymptotic behavior of the solutions
is AdS. In addition, it is observed that as $r$ approaches
infinity, the Kretschmann scalar goes to $\frac{40}{\ell^4}$
($R_{\alpha
\beta \gamma \delta}R^{\alpha \beta \gamma \delta}\rightarrow \frac{40}{%
\ell^4} $) which may confirm the AdS nature of the solutions,
asymptotically. However, in order to have the asymptotically AdS
solutions, one has to check that the asymptotically symmetry group
is the AdS group.

In the following, we are going to find the thermodynamic quantities for the
above solution. For convenience, we report the results of conserved and
thermodynamic quantities per unit dimensionless volume $\Omega_{3}=\frac{V_3%
}{b^3}$, where $V_3 $ is the volume of the constant $t$ and $r$ hypersurface
with the radius $r_+ $.

First, we calculate the entropy of the solution. To do so, we use
the Bekenstein-Hawking area law since we are working in Einstein
gravity. The area of the horizon is given as
\begin{equation}
A= \int d^3 x \sqrt{-g} \bigg\vert_{r=r_+, t=cte}=\frac{V_3}{b^3} r_+^3,
\end{equation}
hence the entropy per unit volume $\Omega_{3}$ is given by
\begin{equation}
S=\frac{A}{4}=\frac{r_{+}^3}{4} .
\end{equation}

In order to calculate the temperature, one can use the surface gravity
formalism or analytical continuation of metric with its regularity at the
horizon. For the metric of the form (\ref{metric}), both the mentioned
methods lead to the following Hawking temperature
\begin{eqnarray}  \label{tem}
T&&=\frac{1}{4 \pi} \frac{d f(r)}{dr} \big \vert _{r=r_+} \\
&&=\frac{1}{2 \pi r_+^3 } \left( \frac{2 r_+^4}{\ell^2} +\frac{1}{2} \mathtt{%
C}_1 r_+^3 + \mathtt{C}_2 r_+^2+ \mathtt{C}_3 r_+ - e^2 \right).
\nonumber
\end{eqnarray}

To find the ADM mass of the black brane in $d$-dimensions, we read the term $%
\frac{-16 \pi G_{d}}{(d-2) \Omega_{d-2}} \frac{M}{r^{d-3}}$ in the
blackening factor (\ref{black}). The volume of the $(d-2)$-dimensional unit
hypersurface, in our case, is $\Omega_3=\frac{V_3}{b^3} $ and we set $%
G_{5}=1 $. So the mass term per unit volume $\Omega_{3}$ is
\begin{equation}  \label{mass}
M=\frac{3}{16 \pi } \left( \frac{r_+^4}{\ell^2}+\frac{ \mathtt{C}_1}{3}
r_+^3 + \mathtt{C}_2 r_+^2 +2 \mathtt{C}_3 r_+ -2 e^2 {\ln (\frac{r_+}{L}})
\right),
\end{equation}
where ``$L$'' is an arbitrary length constant which is considered to have a
dimensionless argument of logarithmic term.

It is worthwhile to note that the definition of mass used here is in
agreement with the first law of black hole thermodynamics
\begin{equation}
T=\Big(\frac{\partial M}{\partial S}\Big) \Big \vert_{\ell, c_i, e}.
\end{equation}


\section{Critical behavior and van der Waals phase transition}\label{sec3}

In this section, we discuss the critical behavior of the introduced black
brane solution of Eqs. (\ref{metric}) and (\ref{black}) in the extended
phase space. To work in the extended phase space, we consider the
cosmological constant as a thermodynamic pressure defined as
\begin{equation}  \label{pressure}
P=-\frac{\Lambda}{8 \pi}=\frac{3}{4 \pi \ell^2}.
\end{equation}

In this regard, we first find the equation of state by substituting pressure
(\ref{pressure}) in the relation of temperature (\ref{tem}) as follows
\begin{equation}
P=\frac{3}{8 \pi} \left( \frac{e^2}{ r_+^4}-\frac{ \mathtt{C}_3}{r_+^3}-
\frac{ \mathtt{C}_2}{r_+^2}-\frac{ \mathtt{C}_1}{2 r_+}+2 \pi \frac{T}{r_+}
\right).
\end{equation}

It is worth mentioning that, here, the equation of state is $P=P(T,r_{+})$
rather than $P=P(T,v)$. Such situation comes from the fact that the specific
volume ($v$) is related to the event horizon as $v = \frac{{4{r_ + }\ell _{%
\mathrm{P}}^{{3}}}}{{3}}$ (in $5$-dimensions), and therefore, criticality in
$P-v$ diagram is equivalent to the criticality in $P-r_{+}$ plot. From this
equation, we see that it is possible to define an effective (shifted)
temperature $\mathcal{T} $ as below
\begin{equation}
\mathcal{T}=T-\frac{\mathtt{C}_1}{4 \pi}.
\end{equation}

To find the critical point in $P-r_+$ diagram, we attend the inflection
point which is presented by these two equations
\begin{equation}
\Big(\frac{\partial P}{\partial r_+}\Big) \Big \vert_ \mathcal{T}=0,~~~~~%
\Big(\frac{\partial^2 P}{\partial r_+^2}\Big) \Big \vert_ \mathcal{T}=0.
\end{equation}

Solving these equations simultaneously, the critical quantities are obtained
as
\begin{eqnarray}  \label{criticalQ}
&&r_c=- \frac{3 \mathtt{C}_3 \pm \Theta}{2 \mathtt{C}_2}, \\
&& \mathcal{T}_{c}=- \frac{2 \left(9 \mathtt{C}_3^2\pm3 \mathtt{C}_3 \Theta
+16 \mathtt{C}_2 e^2\right) \mathtt{C}_2^2}{\pi (3 \mathtt{C}_3 \pm \Theta
)^3},  \nonumber \\
&&P_c=\frac{ \left(9 \mathtt{C}_3^2\pm 3 \mathtt{C}_3 \Theta +18 \mathtt{C}%
_2 e^2\right) \mathtt{C}_2^3}{\pi (3 \mathtt{C}_3 \pm \Theta)^4},
\nonumber
\end{eqnarray}
where $\Theta =\sqrt{9 \mathtt{C}_3^2 +24 \mathtt{C}_2 e^2}$. As
it is obvious, there are two branches for critical quantities,
which are distinguished by the sign behind $\Theta $. For the
upper sign, we can show that there exists no acceptable critical
behavior with real positive values for all the critical
quantities. In other words, in order to have a positive critical
horizon radius, it is necessary to consider negative
$\mathtt{C}_2$
such that $9 \mathtt{C}_3^2 +24 \mathtt{C}_2 e^2 > 0 $ and positive $\mathtt{%
C}_3$. In this case, both the critical temperature and pressure are
negative. Therefore, there is no acceptable parameter value to have physical
critical behavior by the upper sign.

As it is known, the planar black branes in the Einstein-AdS gravity have no
van der Waals phase transition and critical behavior. So the important
concern is that obtaining the criticality of our solution may arise from the
massive term or the Yang-Mills one or both terms. To enlighten such concern,
we consider two below cases and do the criticality process again

\begin{itemize}
\item $m=0 $ to have just Yang-Mills term

\item $e=0 $ to have just massive term
\end{itemize}

After some calculations, one observes that in the first case, in agreement
with the Einstein-AdS gravity, the criticality does not occur. Therefore,
the Yang-Mills term cannot change the existence/absence of criticality. But
the second case is different and the critical quantities are
\begin{equation}
r_c=-\dfrac{3 \mathtt{C}_3}{\mathtt{C}_2},~~\mathcal{T}_c=-\dfrac{ \mathtt{C}%
_2^2}{6 \pi \mathtt{C}_3},~~P_c=\dfrac{ \mathtt{C}_2^3}{72 \pi \mathtt{C}_3^2%
}.
\end{equation}

We see that to have positive critical values, it is necessary to set $%
\mathtt{C}_3<0$ and $\mathtt{C}_2>0 $. Hence, we get the result
that the existence/absence of criticality of the planar black
brane solution comes from the massive term. It is also notable
that although the Yang-Mills term does not have the main role in
the existence/absence of criticality, it can affect the values of
possible critical quantities. To study the effects of Yang-Mills
and massive parameters on the critical quantities, two tables are
given. In table \ref{table1}, it is observed that increasing
$``e"$ leads to increasing $r_c $ and decreasing $\mathcal{T}_c $
and $P_c $. While according to table \ref{table2}, we find that by
increasing massive
parameters $\mathtt{C_2} $ and $\mathtt{C_3} $, $r_c $ decreases and $%
\mathcal{T}_c $ and $P_c $ increase. Besides, as you see from Eq.
(\ref{criticalQ}), changing $\mathtt{C}_1 $ does not impress the
critical quantities. It comes from the fact that the role of
$\mathtt{C}_1 $ is only a shift in the temperature, and therefore,
it does not have a direct effect on the critical quantities. So,
we put it aside from the characteristics of the tables.

To more explore the phase transition and its order, we need to calculate
other quantities. In this regard, the Gibbs free energy per unit volume $%
\Omega_{3}$ is as follows
\begin{eqnarray}
G=M-T S &=&\frac{1 }{8 \pi } \left[- \dfrac{2}{3} r_+^4 \pi P
+\frac{1}{2}
\mathtt{C}_2 r_+^2+ 2 \mathtt{C}_3 r_{+}\right.  \nonumber \\
&& \left. + e^2 - 3 e^2 {\ln (\frac{r_+}{L}}) \right],
\end{eqnarray}
and the specific heat per unit volume $\Omega_{3}$ is given by
\begin{eqnarray}\label{specificheat}
&&C_{P}= T \frac{\partial S}{\partial T} \bigg\vert _{\Lambda, c_i, e}= \\
&&\frac{3}{8 }\frac{ \left(16 P \pi r_+^4+3 (2 \mathtt{C}_3 r_+ +2 \mathtt{C}%
_2 r_+^2 + \mathtt{C}_1 r_+^3) - 6 e^2 \right) r_+^3 }{8 P \pi
r_+^4 -3 (2 \mathtt{C}_3 r_+ + \mathtt{C}_2 r_+^2 ) +9 e^2}.
\nonumber
\end{eqnarray}

\begin{figure}[h!]
\begin{center}
$%
\begin{array}{c}
\includegraphics[width=65 mm]{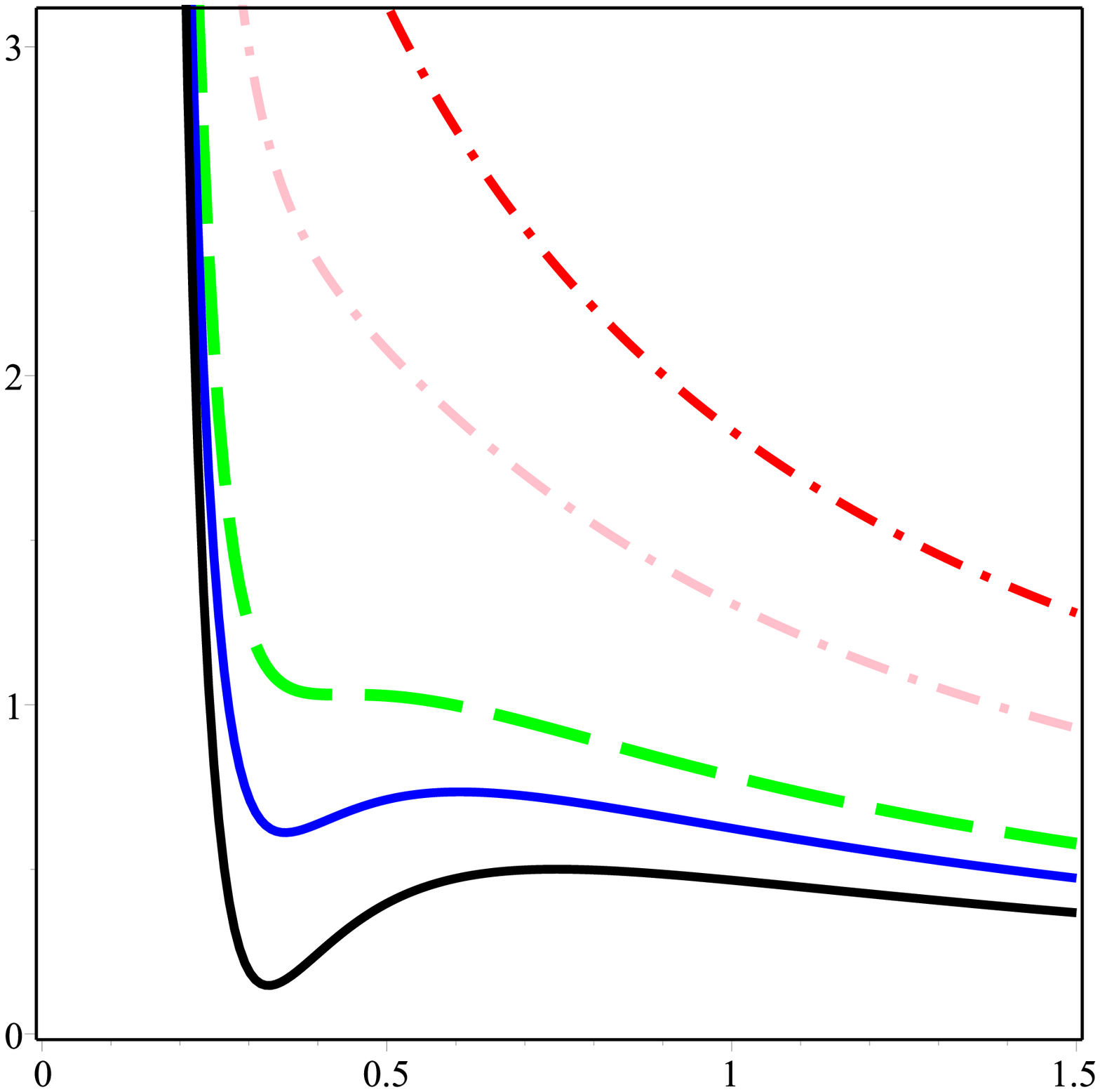} \\
\includegraphics[width=65 mm]{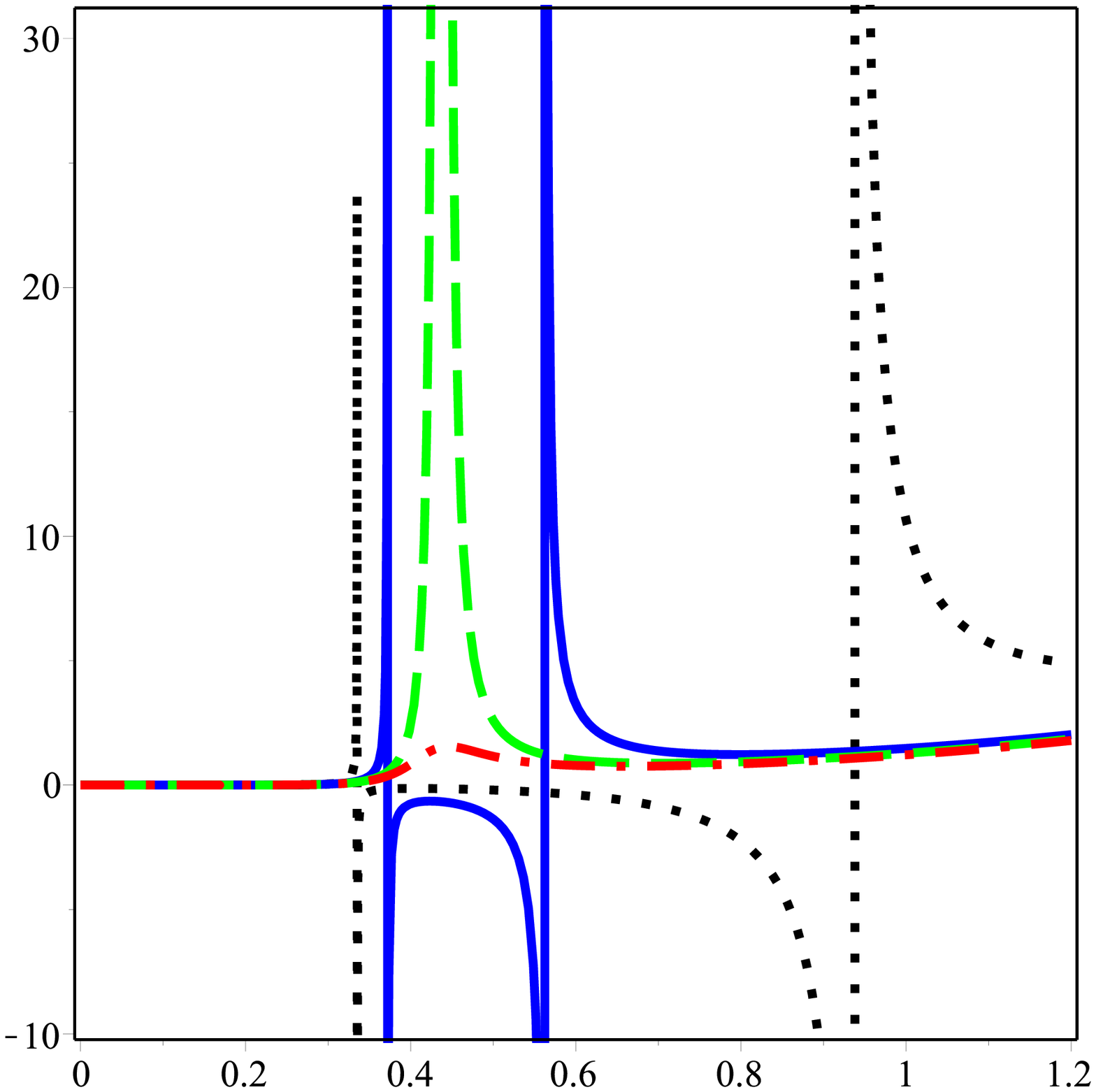} \\
\includegraphics[width=65 mm]{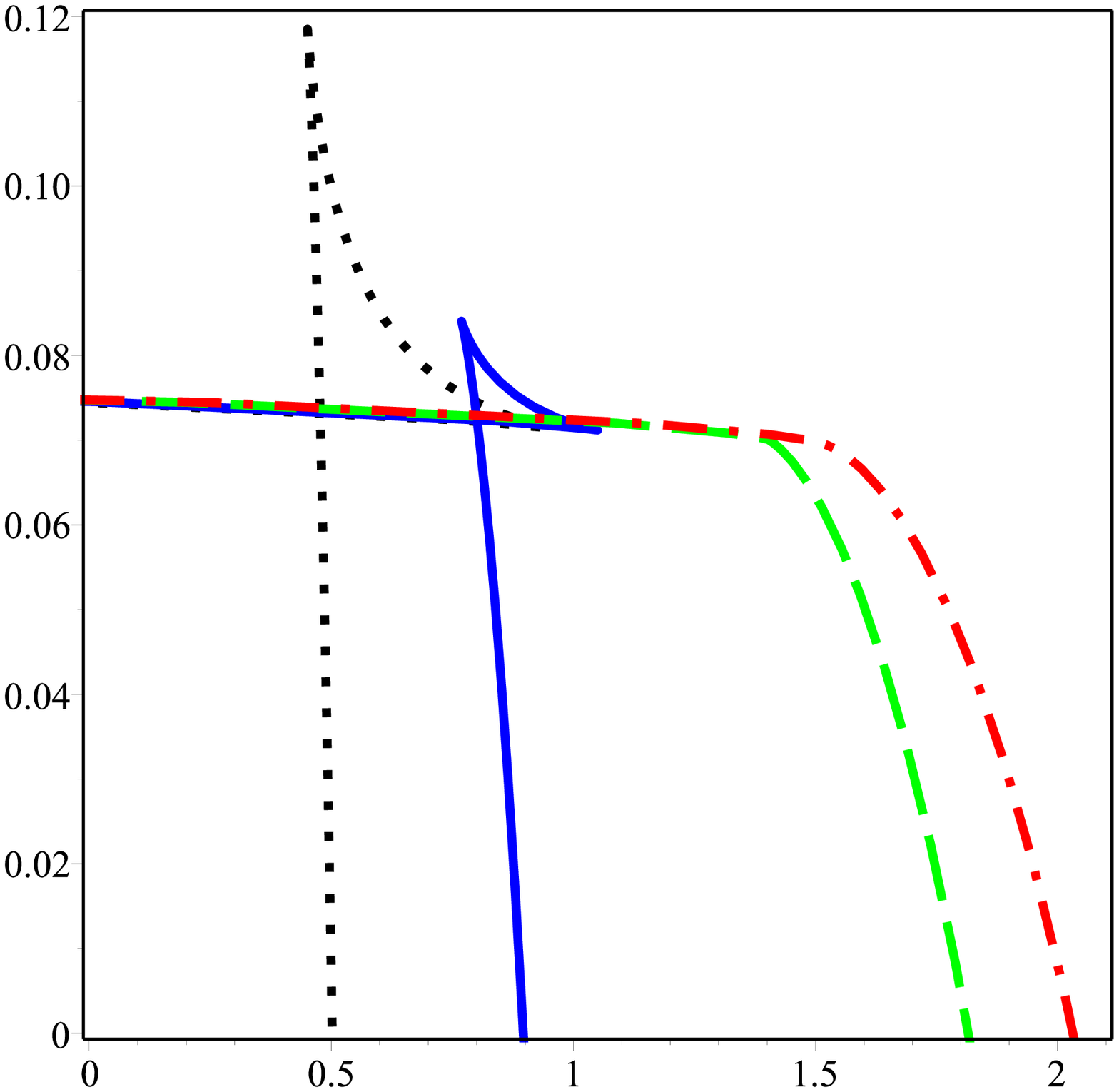}%
\end{array}%
$%
\end{center}
\caption{$P-r_+$ (up), $C_P-r_+$ per unit volume (middle) and $G-T$ per unit
volume (down) diagrams for $e=0.5$, $m=1$, $c_0=c_1=c_2=c_3=1 ( \mathtt{C}%
_i=1)$ and $L=1 $. The green line corresponds to the critical temperature
(up) and the critical pressure (middle and down).}
\label{Fige1}
\end{figure}

In order to explore the van der Waals like phase transition, we
plot some related diagrams in Fig. \ref{Fige1}. In this figure,
the van der Waals like behavior is plotted for the lower sign of
Eq. (\ref{criticalQ}). The critical temperature is associated with
the green dashed line and one can
find the critical point $r_c=0.44 , P_c=1.03 $ for $e=0.5$, $m=1$ and $%
c_0=c_1=c_2=c_3=1 ( \mathtt{C}_i=1)$. According to this figure, one can see
a first order phase transition which is characterized by two divergencies of
the specific heat and the swallow-tail of the Gibbs free energy for $P<P_c $%
. The green line, associated with the critical pressure, exhibits
divergency of the specific heat at the critical point in the
middle panel and discontinuity in the first derivative of the free
energy at the critical point $G_c=0.07 $, $T_c=1.4 $ in down
panel. In the middle panel, the phase transition occurs between
the stable small and large black holes with positive specific
heat, neglecting the middle non-stable states with negative ones.
The two discontinuities in the first derivative of the Gibbs free
energy which are seen in the swallow-tail are the place of
occurrence of divergencies in the specific heat. By increasing the
distance between two divergencies the swallow-tail will be bigger,
which is an indication of the size of non-physical region with
negative specific heat and oscillating part in the $P-r_{+} $
diagram.


\begin{table}[tbp]
\centering
\begin{tabular}{cccc}
\hline\hline
\hspace{0.3cm}$e$\hspace{0.3cm} & \hspace{0.3cm}$r_{c}$ \hspace{0.3cm} &
\hspace{0.3cm} $\mathcal{T}_{c}$\hspace{0.3cm} & \hspace{0.3cm} $P_{c}$%
\hspace{0.3cm} \\ \hline\hline
$0.5$ & $0.4$ & $1.32$ & $1.031$ \\ \hline
$1$ & $1.4$ & $0.24$ & $0.055$ \\ \hline
$1.5$ & $2.5$ & $0.11$ & $0.014$ \\ \hline
$2$ & $3.6$ & $0.07$ & $0.006$ \\ \hline
\end{tabular}%
\caption{Critical values for $\mathtt{C}_2 =1, \mathtt{C}_3 =1 \newline
(m=c_0=c_2=c_3=1)$ with different Yang-Mills parameter.}
\label{table1}
\end{table}

\begin{table}[tbp]
\centering
\begin{tabular}{cccc}
\hline\hline
\hspace{0.3cm}$\mathtt{C}_2 $\hspace{0.3cm} & \hspace{0.3cm}$r_{c}$ \hspace{%
0.3cm} & \hspace{0.3cm} $\mathcal{T}_{c}$\hspace{0.3cm} & \hspace{0.3cm} $%
P_{c}$\hspace{0.3cm} \\ \hline\hline
$0.5$ & $1.6$ & $0.13$ & $0.03$ \\ \hline
$1$ & $1.4$ & $0.24$ & $0.05$ \\ \hline
$1.5$ & $1.2$ & $0.36$ & $0.09$ \\ \hline
$2$ & $1.1$ & $0.50$ & $0.13$ \\ \hline
\end{tabular}
\hspace{1cm} 
\begin{tabular}{cccc}
\hline\hline
\hspace{0.3cm}$\mathtt{C}_3 $\hspace{0.3cm} & \hspace{0.3cm}$r_{c}$ \hspace{%
0.3cm} & \hspace{0.3cm} $\mathcal{T}_{c}$\hspace{0.3cm} & \hspace{0.3cm} $%
P_{c}$\hspace{0.3cm} \\ \hline\hline
$0.5$ & $1.8$ & $0.14$ & $0.02$ \\ \hline
$1$ & $1.4$ & $0.24$ & $0.05$ \\ \hline
$1.5$ & $1.1$ & $0.40$ & $0.12$ \\ \hline
$2$ & $0.9$ & $0.66$ & $0.26$ \\ \hline
\end{tabular}
\\[0pt]
\caption{ up table: critical values for $e=\mathtt{C}_3=1 \newline
(e=m=c_0=c_3=1)$ with different $\mathtt{C}_2 $ parameter.\\[0pt]
down table: critical values for $e= \mathtt{C}_2=1 \newline
(e=m=c_0=c_2=1)$ with different $\mathtt{C}_3 $ parameter.}
\label{table2}
\end{table}

Let us have a look at the local thermal stability of solutions in
different ensembles. Thermal stability of black holes may be
assured in two canonical and grand canonical ensembles by the
positive determinant of the Hessian matrix with respect to
extensive parameters, i.e. $ H_{ij}=\frac{\partial^{2} M}{\partial
x_i
\partial x_j} $. In the canonical ensemble with constant charges,
the Hessian matrix has just one component
\begin{equation*}
H=\frac{\partial^{2} M}{\partial S^2}=\frac{T}{C_p},
\end{equation*}
which besides the necessity for positive temperature, thermal
stability implies $ C_p>0 $. However, in the grand canonical
ensemble, other extensive parameters are allowed to change. By
considering the massive parameters as extensive variables, it
turns out that determinant of the Hessian matrix is zero. To get a
situation in which both ensembles lead to the same thermal
stability condition, we use the freedom in choosing the constant
parameter $ L $ which was introduced in mass term (\ref{mass}) for
making dimensionless argument in the logarithmic function (see
\cite{Mamasani} as an example). We insert $
L=\ell=\sqrt{-6/\Lambda} $ and the Hessian matrix as
\begin{equation}\label{Hessian}
H =
\begin{pmatrix}
\dfrac{\partial^{2} M}{\partial S^2} & \dfrac{\partial^{2} M}{\partial S \partial e} & \dfrac{\partial^{2} M}{\partial S \partial P}  \\ \\
\dfrac{\partial^{2} M}{\partial e\partial S} & \dfrac{\partial^{2} M}{\partial e^2 } & \dfrac{\partial^{2} M}{\partial e \partial P} \\ \\
\dfrac{\partial^{2} M}{\partial P \partial S } &
\dfrac{\partial^{2} M}{\partial P \partial e} &
\dfrac{\partial^{2} M}{\partial P^2}
\end{pmatrix}.
\end{equation}

\begin{figure}[h!]
\begin{center}
\includegraphics[width=70 mm]{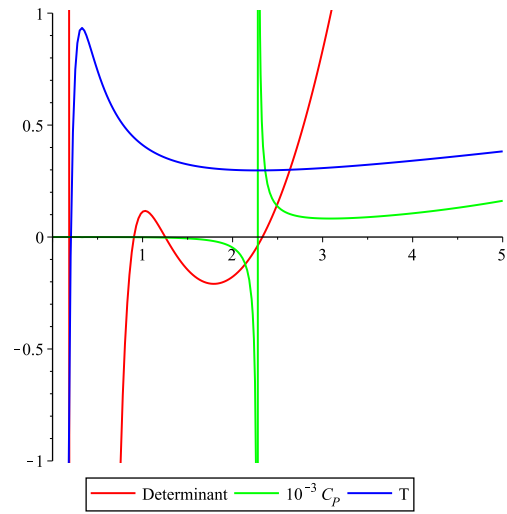}
\end{center}
\caption{Thermal stability in two ensembles with positive
temperature for $e=0.5$, $ \mathtt{C} _i=1$ and $P=\frac{1}{8 \pi}
$.} \label{FigTstability}
\end{figure}

In this way, one can see from Fig. \ref{FigTstability} that
ignoring subtle differences, there is similar region for positive
specific heat (\ref{specificheat}) and positive determinant of
Hessian matrix (\ref{Hessian}).

As we mentioned before, the mass term in (\ref{mass}) satisfies
the first law of thermodynamics which in the extended phase space
can be written as
\begin{equation}  \label{first law}
dM=T dS+V dP+ \sum_{i=1}^{3}\mathcal{C}_i d \mathtt{C}_i+ \mathtt{e}\; d e,
\end{equation}
where the thermodynamic volume per $\Omega_{3}$ is defined as
\begin{equation}
V=\dfrac{\partial M}{\partial P}\bigg\vert_{r_+,c_i,e}=\dfrac{r_+^4}{4},
\end{equation}
and conjugates to the massive parameters and the Yang-Mills coupling
constant per unit volume $\Omega_{3}$ are, respectively,
\begin{equation}
\mathcal{C}_i=\dfrac{\partial M}{\partial c_i}\bigg\vert_{r_+,P,e}=\dfrac{%
i(i+1)}{32 \pi} r_{+}^{4-i},
\end{equation}
and
\begin{equation}
\mathtt{e}=\dfrac{\partial M}{\partial e}\bigg\vert_{r_+,P,c_i}=-\dfrac{3 e}{%
4 \pi} \ln (\frac{r_+}{L}).
\end{equation}

Due to the existence of logarithmic term in the metric function, obtaining
the Smarr formula is not a trivial task. To explore the Smarr formula, we
need to define a new Yang-Mills variable
\begin{equation}
E^{2}=e^{2}\ln (\frac{r_{+}}{L}).  \label{newYM}
\end{equation}%
By substitution of this definition in the mass (\ref{mass}), one obtains
the mass term per unit volume $\Omega _{3}$ as
\begin{equation}
M=\frac{3}{16\pi }\left( \frac{r_{+}^{4}}{\ell ^{2}}+\frac{\mathtt{C}_{1}}{3}%
r_{+}^{3}+\mathtt{C}_{2}r_{+}^{2}+2\mathtt{C}_{3}r_{+}-2E^{2}\right) ,
\label{new mass}
\end{equation}
where the temperature can be calculated as%
\begin{equation}
T=\frac{1}{2\pi r_{+}^{3}}\left( \frac{2r_{+}^{4}}{\ell ^{2}}+\frac{1}{2}%
\mathtt{C}_{1}r_{+}^{3}+\mathtt{C}_{2}r_{+}^{2}+\mathtt{C}_{3}r_{+}\right) .
\label{new T}
\end{equation}

We should note that there is not any effect of Yang-Mills term in
temperature (the last term in Eq. (\ref{tem}) eliminates). It is
notable that although it seems $E$ depends on $r_{+}$, such a
temperature that can be calculated from both the surface gravity
and the first law may confirm that we should interpret $E$ as an
independent thermodynamic quantity. In
other words, one may adjust ``$L$" such a way that the quantity $(\frac{r_{+}%
}{L})$ is considered as a constant. By doing dimensional analysis (for
example with the help of Eq. (\ref{new mass})), the Smarr formula has been
verified
\begin{equation}
2M=3TS-2VP-\mathcal{C}_{1}\mathtt{C}_{1}+\mathcal{C}_{3}\mathtt{C}_{3}+%
\mathcal{E}E,  \label{Smarr}
\end{equation}%
where now we have $\mathcal{E}=\dfrac{\partial M}{\partial E}\bigg\vert%
_{r_{+},P,c_{i}}=-\dfrac{3E}{4\pi }$. Please note that the massive
parameter $\mathtt{C}_{2}$ does not appear in the Smarr formula
since it has no scaling and  so it is not a thermodynamic
variable. Moreover, for this reason, we can omit this parameter
from the first law (\ref{first law}). The role of the new quantity
$``E"$ in the Smarr formula suggests that one may consider $``E"$
as a thermodynamic variable instead of $``e"$ with the following
explicit form
\begin{equation}
dM=TdS+VdP+\mathcal{C}_{1}d\mathtt{C}_{1}+\mathcal{C}_{3}d\mathtt{C}_{3}+%
\mathcal{E}dE.  \label{new first law}
\end{equation}

\section{PHOTON SPHERE AND SHADOW}\label{sec 4}
 \begin{figure*}[!htb]
\centering
    \includegraphics[width=0.38\linewidth]{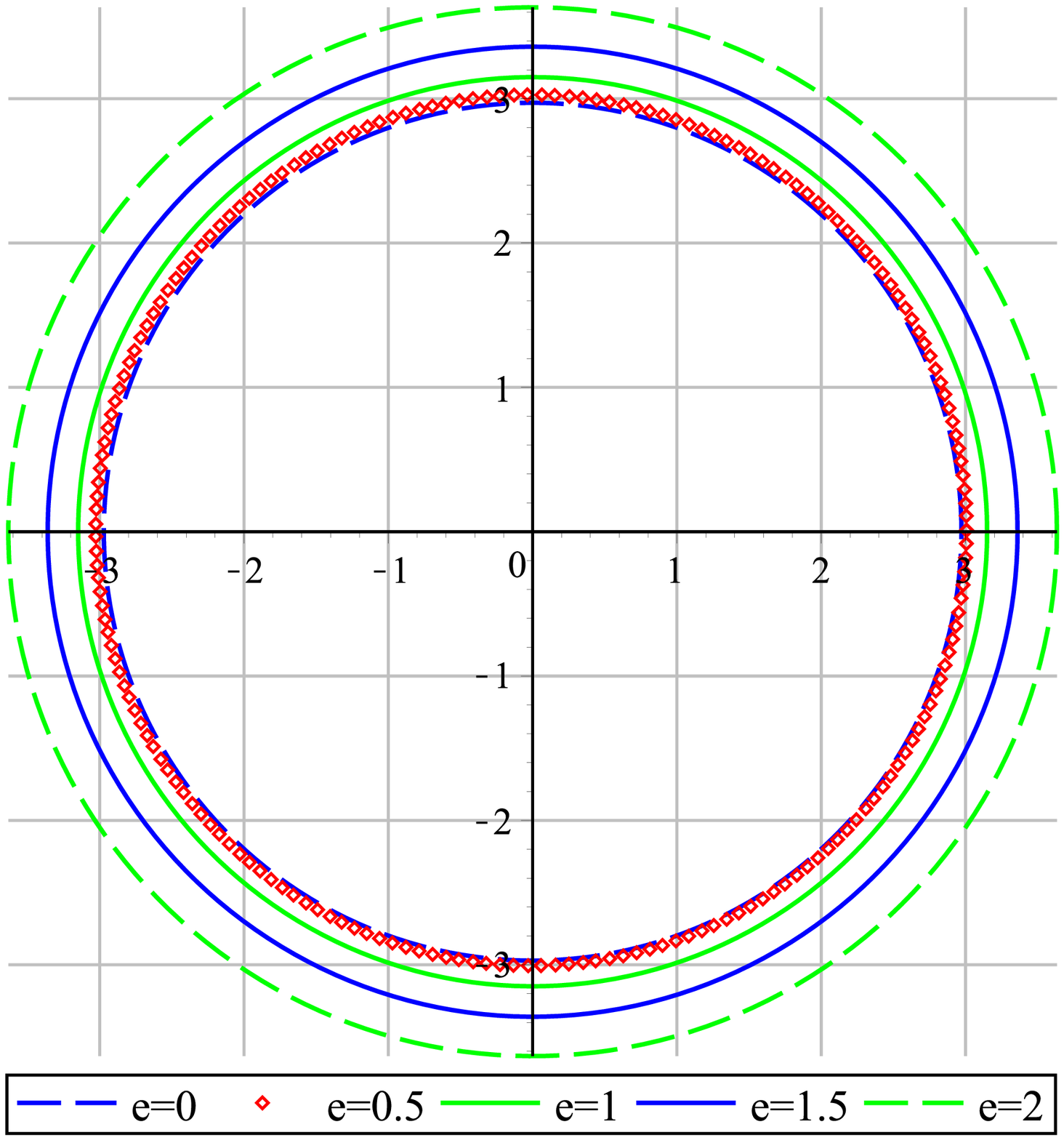}\hfil
    \includegraphics[width=0.42\linewidth]{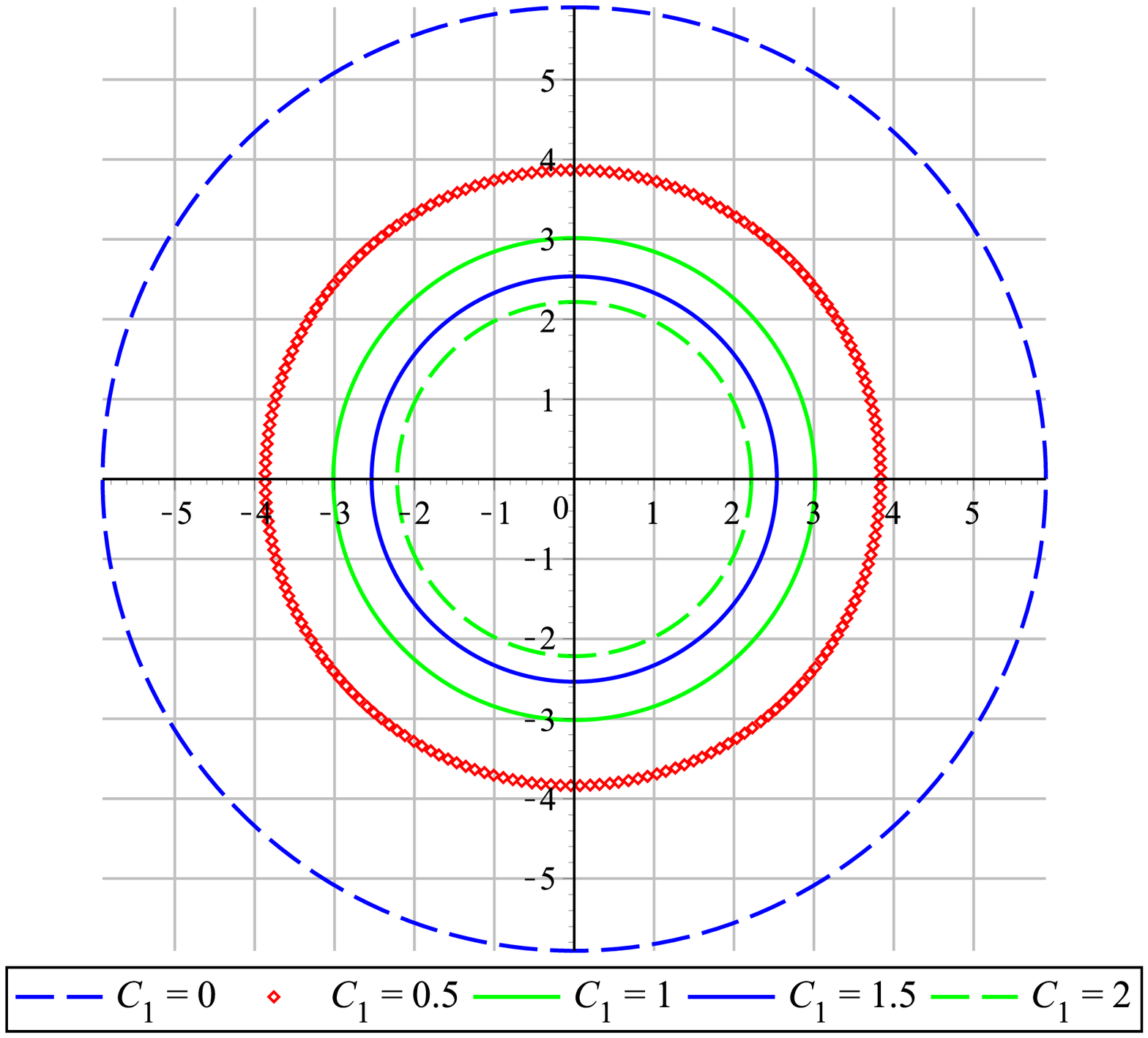}\par\medskip
    \includegraphics[width=0.47\linewidth]{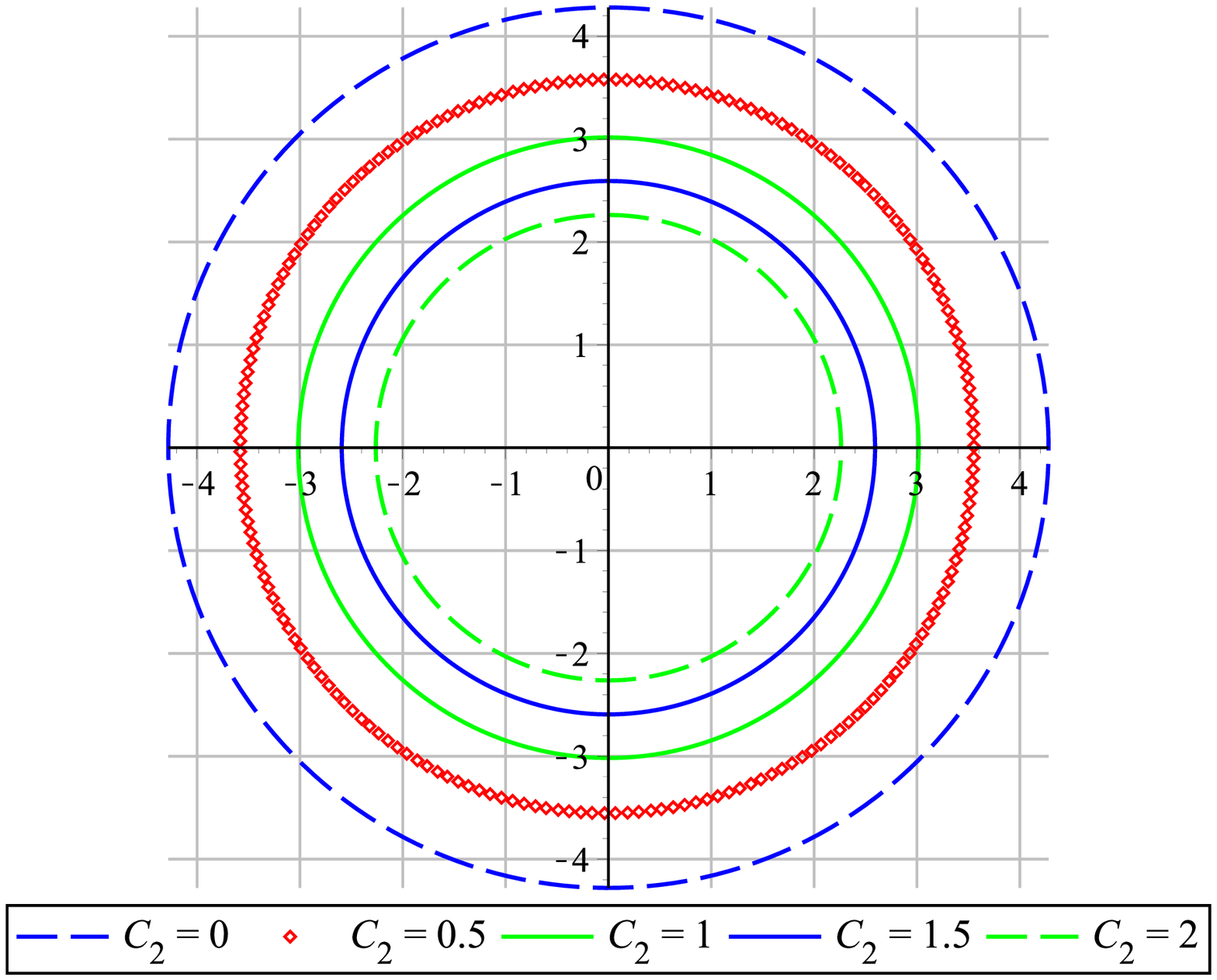}\hfil
    \includegraphics[width=0.42\linewidth]{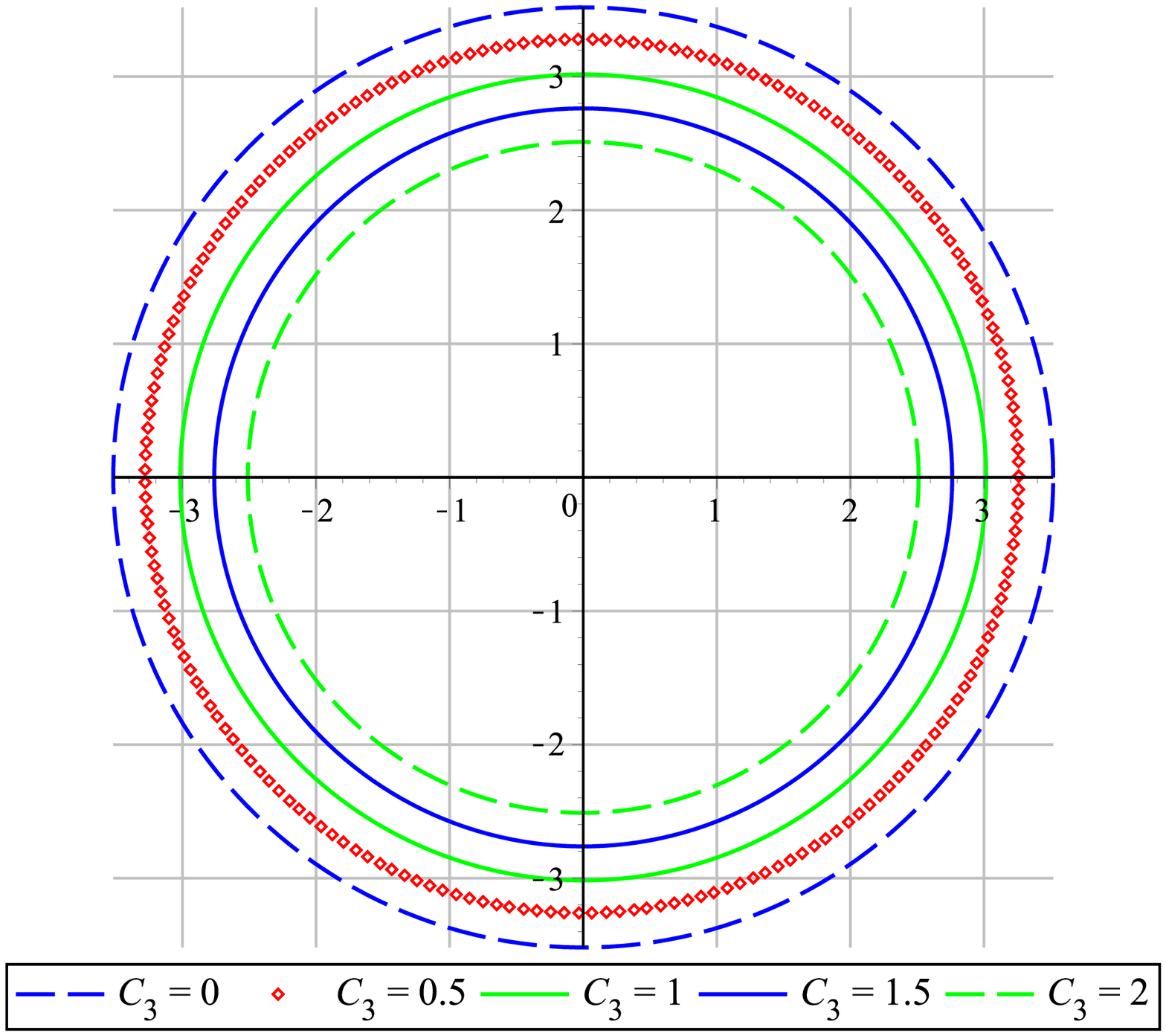}
\caption{Black hole shadow in the Celestial plane ($\alpha-\beta$)
for varying $ e $ with $\mathtt{C}_i=1$ (up-left panel), varying
$\mathtt{C}_1 $ with $\mathtt{C}_2=\mathtt{C}_3=1$ and $ e=0.5 $
(up-right panel), varying $\mathtt{C}_2 $ with
$\mathtt{C}_1=\mathtt{C}_3=1$ and $ e=0.5 $ (down-left panel) and
varying $\mathtt{C}_3 $ with $\mathtt{C}_1=\mathtt{C}_2=1$ and $
e=0.5 $ (down-right panel). We set $ M=L=\ell=1 $.} \label{shadow}
\end{figure*}
It is quite clear that the strong gravitational field of black
holes enforces propagation of light along curved lines such that
the spherical light paths may arise (for example see
\cite{Panpanich:2019mll}). If these photon orbits being unstable,
then photons can scape to infinity and project on the Celestial
sphere by a distant observer. Here, we study this phenomenon for
the black hole solutions under consideration. The unstable
critical photon orbit is known as the ``photon sphere" and one may
look for its radius depending on various parameters.

Considering the metric, Eqs. (\ref{metric}) and (\ref{black}), we
find that the coefficients are independent of $ ``t" $ and
$``x_i"$. So, these symmetries imply the constants of motion $``-
E"$ and $ ``L_{i}" $ for the time and  $ x_i ~(i=1,2,3) $
directions, respectively. Now, we assume a test particle with rest
mass $``m_0" $ moving around the black hole on a curve with an
affine parameter $\lambda$. First, we should find the geodesic
equations of motion using the Lagrangian
\begin{equation}
\mathcal{L}=\frac{1}{2} g_{\mu \nu} \dot{x}^{\mu}\dot{x}^{\nu};~~~\dot{x}^{\mu}=\frac{d x^{\mu}}{d \lambda}.
\end{equation}

Substituting the metric from Eq. (\ref{metric}), we find
\begin{equation}
\mathcal{L}=- \frac{f(r)}{2} \left(\frac{d t}{d
\lambda}\right)^2+\frac{1}{2 f(r)} \left(\frac{d r}{d
\lambda}\right)^2+\frac{r^2}{2 b^2} \delta _{ij} \frac{d x^i}{d
\lambda} \frac{d x^j}{d \lambda}.
\end{equation}

Now, reminding the relations $ E=-\frac{\partial
\mathcal{L}}{\partial\dot{t}} $ and $  L^{i}=\frac{\partial
\mathcal{L}}{\partial\dot{x_i}} $, we can obtain the geodesic
equations of motion as
\begin{equation}
\frac{d t}{d \lambda}=\frac{E}{f(r)},~~~~\frac{d x^{i}}{d
\lambda}=\frac{b^2}{r^2} L^{i}.
\end{equation}

The equation of motion in the radial direction can be derived from
the Hamilton-Jacobi equation
\begin{equation}
S=\frac{1}{2} m_0^2 \lambda - E t+L_1 x_1+L_2 x_2+L_3 x_3+S_r (r),
\end{equation}
where $ S_r (r) $ is a function of $ r $. This equation allows us to find
\begin{equation}
f(r) (\frac{\partial S_r}{\partial
r})^2=-m_0^2+\frac{E^2}{f(r)}-\frac{b^2}{r^2} L^{2},
\end{equation}
in which $L^{2}=L_1^2+L_2^2+L_3^2$. With the help of $
p_r=\frac{\partial \mathcal{L} }{\partial \dot r}= \frac{\partial
S }{\partial r}$, the null geodesic equation with $ m_0=0 $ is
obtained in the following form
\begin{eqnarray}
&&\left(\frac{d r}{d \lambda}\right)^2+V_{eff}(r)=0,\\
&&V_{eff}(r)\equiv -E^2+\frac{b^2}{r^2} f(r) L^{2}.
\end{eqnarray}

To have the spherical geodesics, we apply two conditions
\begin{equation*}
 V_{eff}(r)\bigg\vert_{r=r_p}=0,~~~\dfrac{\partial V_{eff}(r) }{\partial r}\bigg\vert_{r=r_p}=0,
\end{equation*}
where $ r_p $ is the radius of photon orbit. The first condition
results to
\begin{equation}
\dfrac{r_p^2}{b^2 f(r_p)}=\dfrac{L^2}{E^2},
\end{equation}
and the second condition gives
\begin{equation}
-2 f(r_p)+ r_p f' (r_p)=0.
\end{equation}

Substituting the metric function (\ref{black}) in the recent
relation, we find
\begin{equation}\label{PSeq}
-\dfrac{1}{3} \mathtt{C}_1 r_p^3 -2 \mathtt{C}_2   r_p^2-6
\mathtt{C}_3 r_p+\dfrac{64 \pi}{3} M+8 e^2  \ln(\dfrac{r_p}{L})-2
e^2=0.
\end{equation}

In general, analytical solution of this equation for $r_{p}$ is
not a trivial task. However, for some special cases, one can find
an analytical relation for the radius of photon sphere, as follow
\begin{itemize}
\item $ e=0 $
\begin{equation}
r_p=\dfrac{\eta ^2-2 \eta  \mathtt{C}_2 -6 \mathtt{C}_1
\mathtt{C}_3 +4 \mathtt{C}_2^2}{\mathtt{C}_1 \eta},
\end{equation}
where
\begin{eqnarray*}
\eta &=& (32 M \pi  \mathtt{C}_1^2 + 18  \mathtt{C}_1 \mathtt{C}_2
\mathtt{C}_3 -8  \mathtt{C}_2^3 \pm 2 \gamma^{\frac{1}{2}}
\mathtt{C}_1)^{\frac{1}{3}}, \\
\gamma&=&32 \pi M (8 \pi M  \mathtt{C}_1^2 +9 \mathtt{C}_1
\mathtt{C}_2  \mathtt{C}_3 -4 \mathtt{C}_2^3)+ \\
&& 27 \mathtt{C}_3^2( 2 \mathtt{C}_1 \mathtt{C}_3 -
\mathtt{C}_2^2).
\end{eqnarray*}

\item $ \mathtt{C}_i$'s$=0 $
\begin{equation}
r_p=L ~\exp \left(\frac{1}{4}-\frac{8 \pi M}{3 e^2}\right) .
\end{equation}

\item If one of the $\mathtt{C}_i$'s is non zero, then $r_p$ is found in terms of the Lambert W function.
\end{itemize}
\begin{table}[!htb]
\centering
\begin{tabular}{cccccc}
 \footnotesize $\mathtt{C}_1$  \hspace{0.3cm} & \hspace{0.3cm}$0$ \hspace{0.3cm} &
\hspace{0.3cm} $0.5$\hspace{0.3cm} & \hspace{0.3cm} $1$\hspace{0.3cm} & \hspace{0.3cm}$1.5$\hspace{0.3cm} & \hspace{0.3cm}$2$ %
\hspace{0.3cm}  \\ \hline\hline
$ r_p (e=0.5 $, $\mathtt{C}_2=\mathtt{C}_3=1$) & $4.58$ & $ 4.09 $ & $3.78$ & $3.55$ & $ 3.38 $\\ \hline
\\
 \footnotesize $\mathtt{C}_2$  \hspace{0.3cm} & \hspace{0.3cm}$0$ \hspace{0.3cm} &
\hspace{0.3cm} $0.5$\hspace{0.3cm} & \hspace{0.3cm} $1$\hspace{0.3cm} & \hspace{0.3cm}$1.5$\hspace{0.3cm} & \hspace{0.3cm}$2$ %
\hspace{0.3cm}  \\ \hline\hline
$ r_p (e=0.5 $, $\mathtt{C}_1=\mathtt{C}_3=1)$ & $4.94$ & $4.26$ & $3.78$ & $3.42$& $3.14$\\ \hline
\\
 \footnotesize $\mathtt{C}_3$  \hspace{0.3cm} & \hspace{0.3cm}$0$ \hspace{0.3cm} &
\hspace{0.3cm} $0.5$\hspace{0.3cm} & \hspace{0.3cm} $1$\hspace{0.3cm} & \hspace{0.3cm}$1.5$\hspace{0.3cm} & \hspace{0.3cm}$2$ %
\hspace{0.3cm}  \\ \hline\hline
$ r_p (e=0.5 $, $\mathtt{C}_1=\mathtt{C}_2=1)$ & $4.46$ & $4.11$ & $3.78$ & $3.46$& $3.17$\\ \hline
\\
 \footnotesize $e$  \hspace{0.3cm} & \hspace{0.3cm}$0$ \hspace{0.3cm} &
\hspace{0.3cm} $0.5$\hspace{0.3cm} & \hspace{0.3cm} $1$\hspace{0.3cm} & \hspace{0.3cm}$1.5$\hspace{0.3cm} & \hspace{0.3cm}$2$ %
\hspace{0.3cm}  \\ \hline\hline
$ r_p (\mathtt{C}_1=\mathtt{C}_2=\mathtt{C}_3=1) $ & $3.72$ & $3.78$ & $3.97$ & $4.29$& $4.74$\\ \hline
\end{tabular}%
\caption{Photon sphere radius for variation of massive and
Yang-Mills parameters for $M =1$ and $L =1$.} \label{table3}
\end{table}
It is possible to numerically show that there are two positive
roots for the Eq. (\ref{PSeq}), which are larger than the event
horizon radius and therefore we have two small and large spherical
light orbits. To know which one is stable with respect to the
radial perturbations, we rely on the sign of the second derivative
of $V_{eff}(r)$, where $ V_{eff}^{\prime\prime}(r_p)>0 $ indicates
stable orbits and $ V_{eff}^{\prime\prime}(r_p)<0 $ stands for
unstable ones. The radius of the unstable photon sphere (the
larger photon orbit) for some parameters is listed in table
\ref{table3}, thereby we understand the behavior of $ r_p $ under
variation of the massive and Yang-Mills parameters. We see that
the radius decreases with increasing the massive parameters while
keeping the Yang-Mills parameter constant. Furthermore, with
constant massive parameters, we observe that the bigger Yang-Mills
parameter is, the larger photon sphere we have.

To discuss the black hole shadow, we consider one luminous source
behind the black hole with a strong gravitational field to lens
the radiation coming from the source. We change the Cartesian
coordinates of the horizon to the spherical ones to have more
straightforward calculations. One can follow the calculations of
Ref. \cite{Das:2019sty} for the shadow radius of black holes with
spherically symmetric metric. For this purpose, we go to the
Celestial coordinates which are introduced with $ \alpha $ as the
perpendicular distance of the shadow from the axis of symmetry and
$ \beta $ as the apparent perpendicular distance of the shadow
from its projection on the equatorial plane. In this way, we get
an equation representing a circle of radius $ R_s $ in the
Celestial plane $ \alpha-\beta $, given by
\begin{equation}
\alpha^{2}+\beta^{2}=R_s^2= \dfrac{\dfrac{r_p^2} {f(r_p)}}
{1-\dfrac{r_p^2}{f(r_p)}\dfrac{f(r_o)}{r_o^2}},
\end{equation}
where subscripts $``p"$ and $``o" $ denote the photon sphere and
observer, respectively. Considering Eq. (\ref{black}), one finds $
\frac{f(r_o)}{r_o^2}\rightarrow \frac{1}{\ell^2} $ (asymptotic AdS
spacetime) for the limit $ r_o \rightarrow \infty $ for the
distant observer. So the radius of shadow calculating by a distant
observer is given by
\begin{equation}
R_s=\frac{\ell}{\sqrt{\frac{f(r_p) \ell^2   }{r_p^2}-1}},
\end{equation}
in which for the asymptotically flat black hole, it reduces to $
R_s=\frac{r_p}{\sqrt{f (r_p)}} $.

In Fig. \ref{shadow}, we plot the black hole shadow for different
Yang-Mills and massive parameters. We find that the shadow size
shrinks with decreasing $``e" $ (increasing $ \mathtt{C}_i $),
which is analogous with the obtained results for the variation of
photon sphere radius. Furthermore, we see that variation of $``e
"$ has weaker effect on the shadow size than the massive
parameters, while more significant effect is occurred for
$\mathtt{C}_1$.
\section{Quasinormal Modes}\label{sec5}

\begin{figure*}[!htb]
\centering
\includegraphics[width=65 mm]{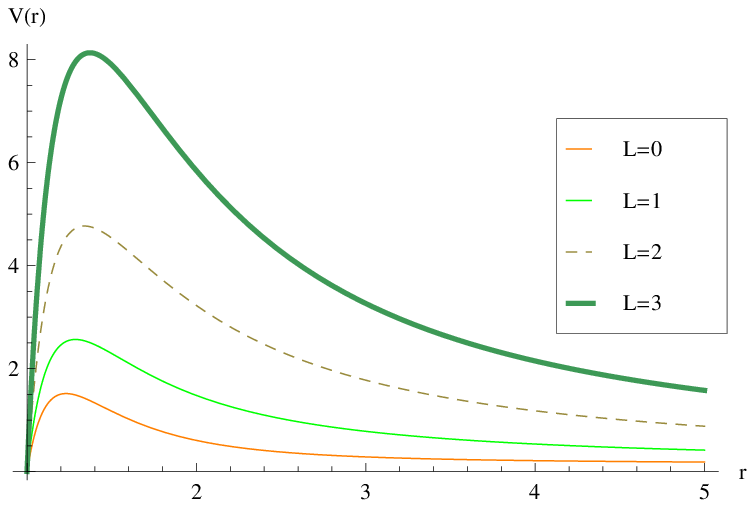} \hfil
\includegraphics[width=65 mm]{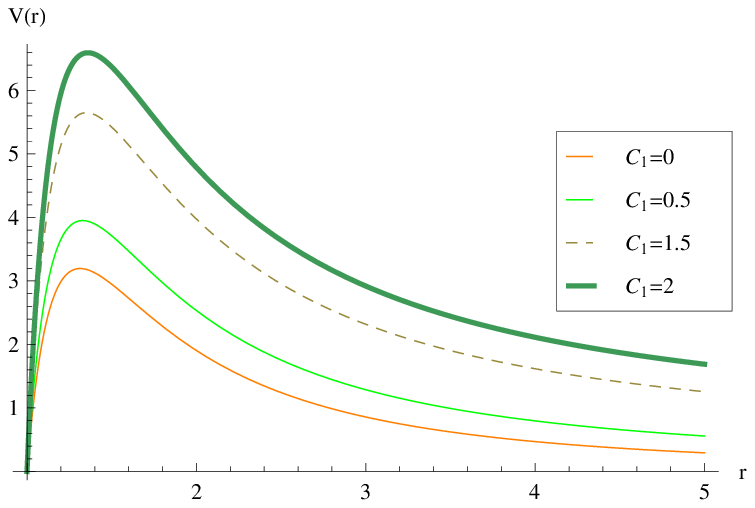} \par\medskip
\includegraphics[width=65 mm]{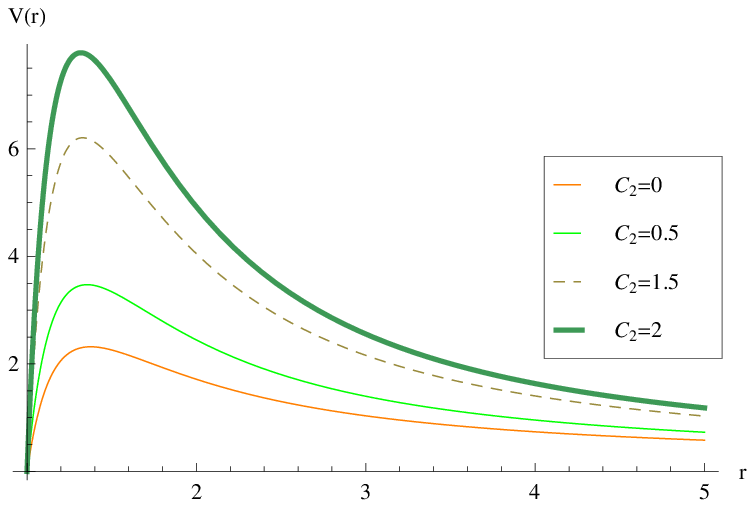} \hfil
\includegraphics[width=65 mm]{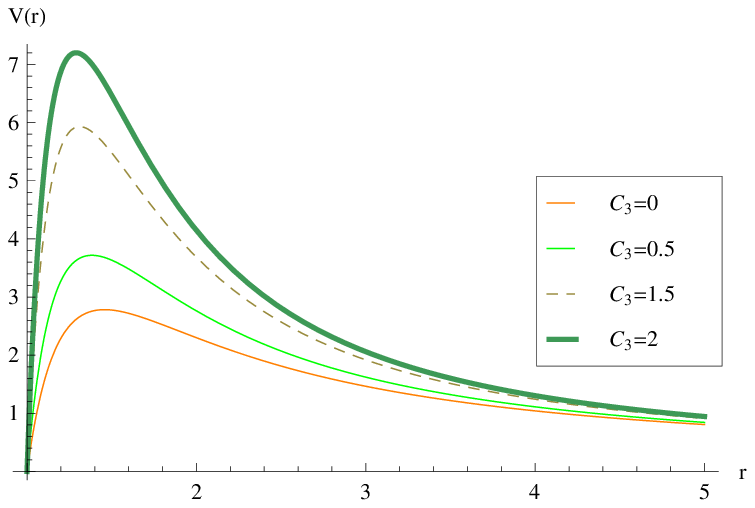} \par\medskip
\includegraphics[width=65 mm]{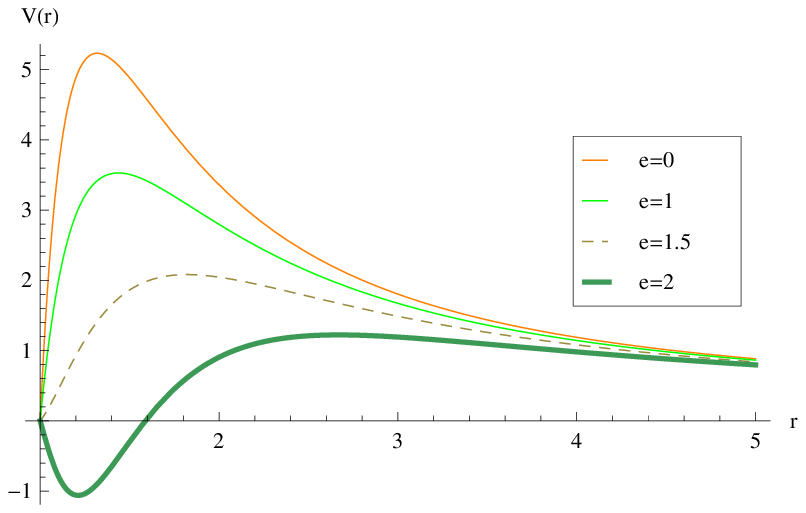}
\caption{Effective potential for $\mathtt{C}_1 =\mathtt{C}_2=\mathtt{C}_3=1 $, $ e $=0.5  with different $ L $ (up-left panel), $\mathtt{C}_2=\mathtt{C}_3=1$, $ e=0.5 $ with $ L =2 $ (up-right panel),  $\mathtt{C}_1=\mathtt{C}_3=1$, $ e=0.5 $ with $ L =2 $  (middle-left panel), $\mathtt{C}_1=\mathtt{C}_2=1$, $ e=0.5 $ with $ L =2 $  (middle-right panel), $\mathtt{C}_1 =\mathtt{C}_2=\mathtt{C}_3=1 $ with $ L =2 $  (down panel).}
\label{potential}
\end{figure*}
\begin{figure*}[!htb]
    \centering
    \includegraphics[width=65 mm]{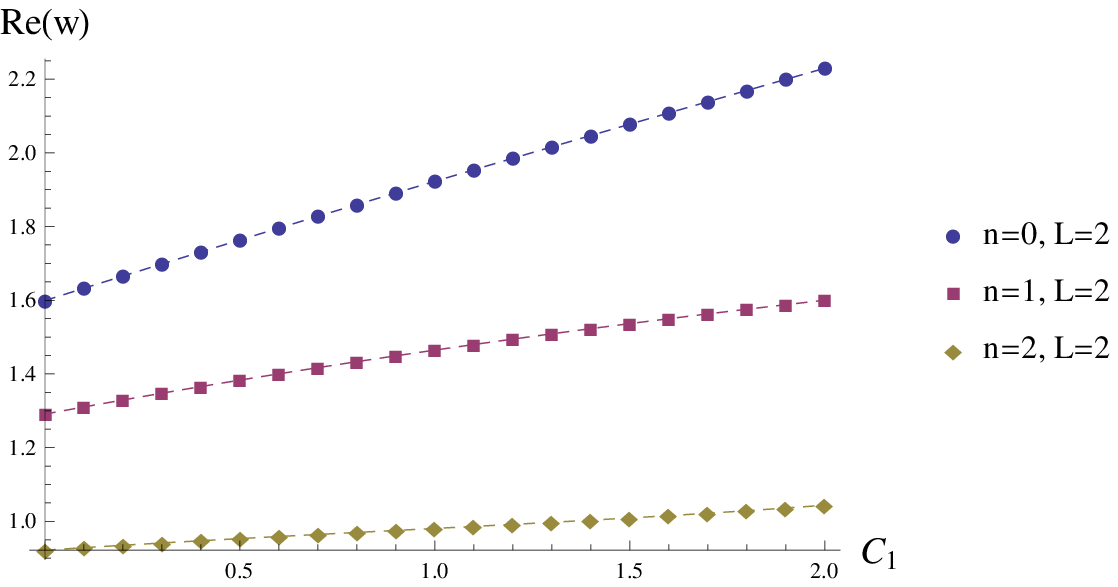} \hfil
\includegraphics[width=65 mm]{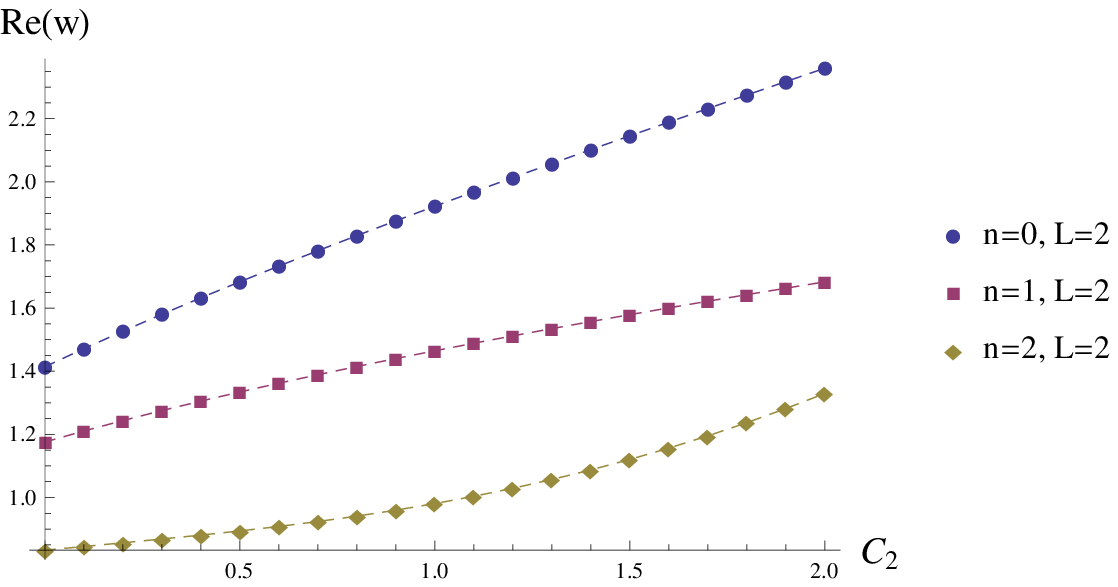} \par\medskip
\includegraphics[width=65 mm]{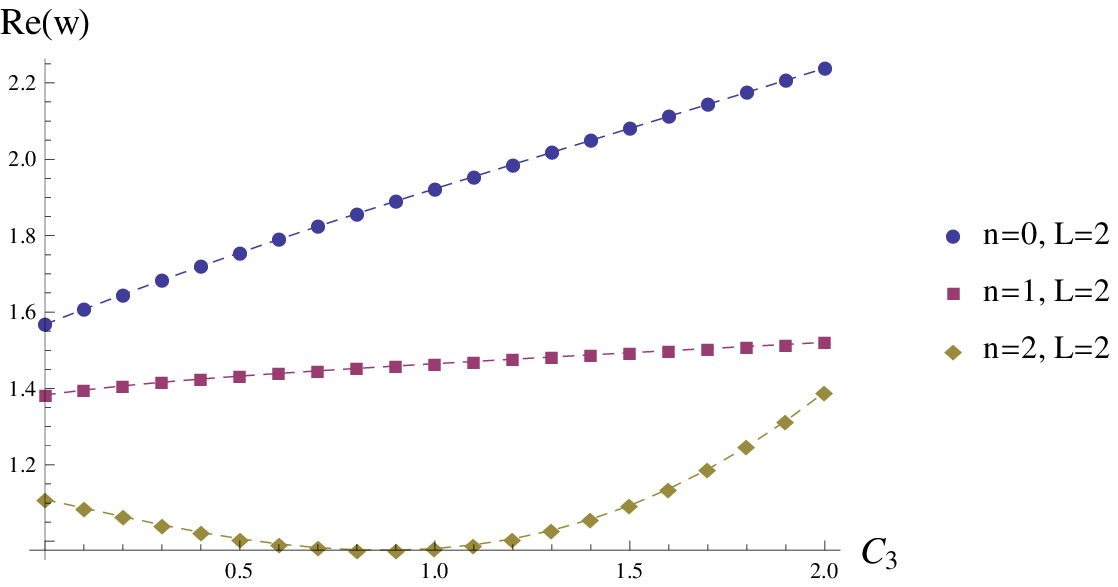} \hfil
\includegraphics[width=65 mm]{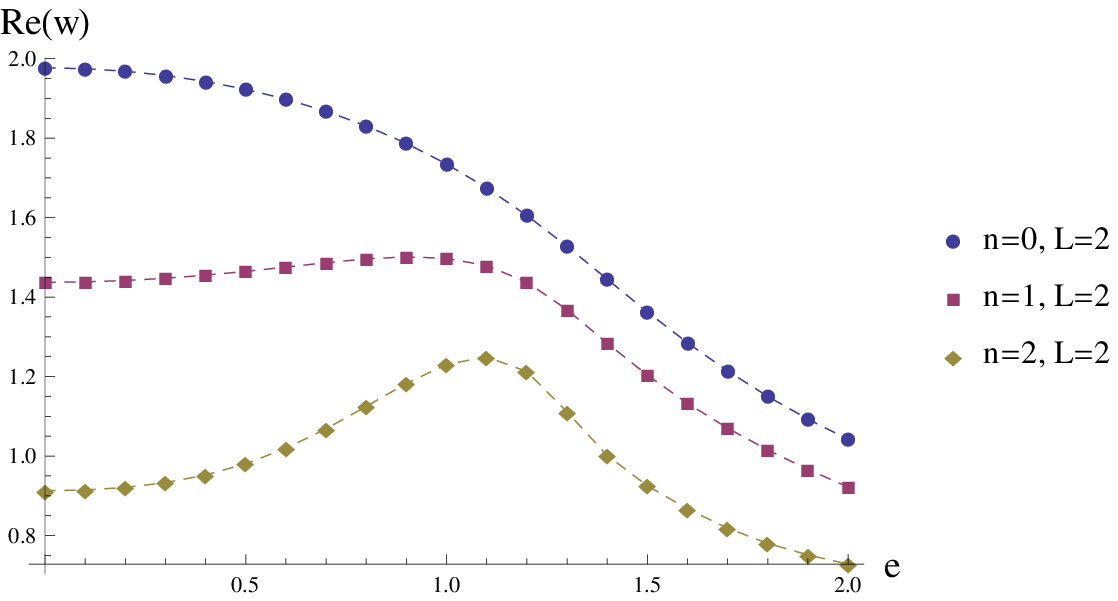}
    \caption{Real part of scalar quasinormal modes for $\mathtt{C}_2=\mathtt{C}_3=1 $, $ e $=0.5 (up-left panel), $\mathtt{C}_1=\mathtt{C}_3=1 $, $ e $=0.5 (up-right panel), $\mathtt{C}_1 =\mathtt{C}_2=1 $, $ e $=0.5 (down-left panel), $\mathtt{C}_1 =\mathtt{C}_2=\mathtt{C}_3=1 $ (down-right panel).}
\label{real}
\end{figure*}
\begin{figure*}[!htb]
    \centering
 \includegraphics[width=65 mm]{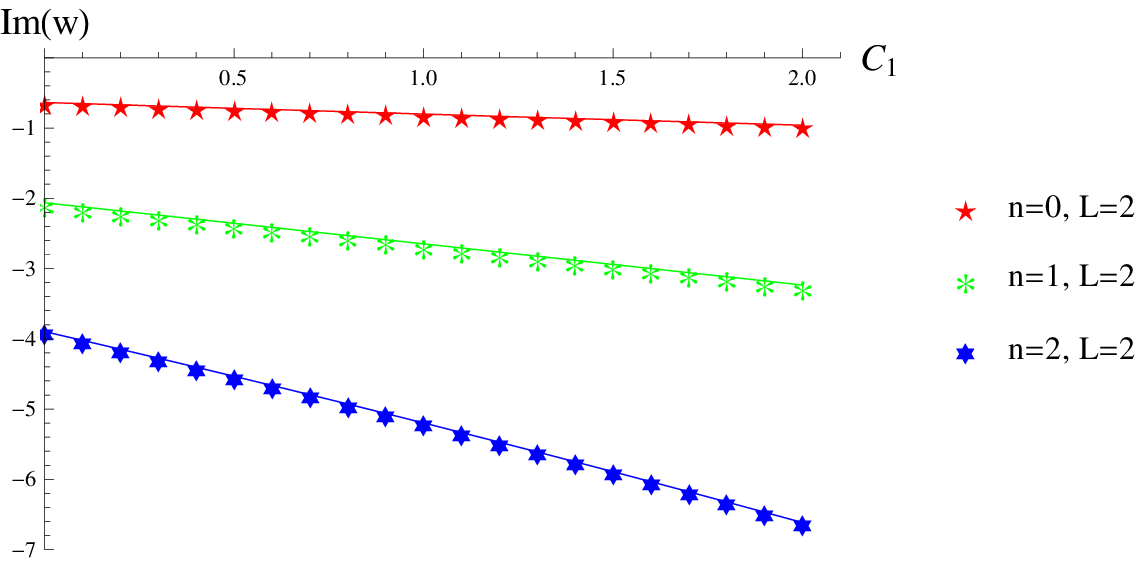} \hfil
\includegraphics[width=65 mm]{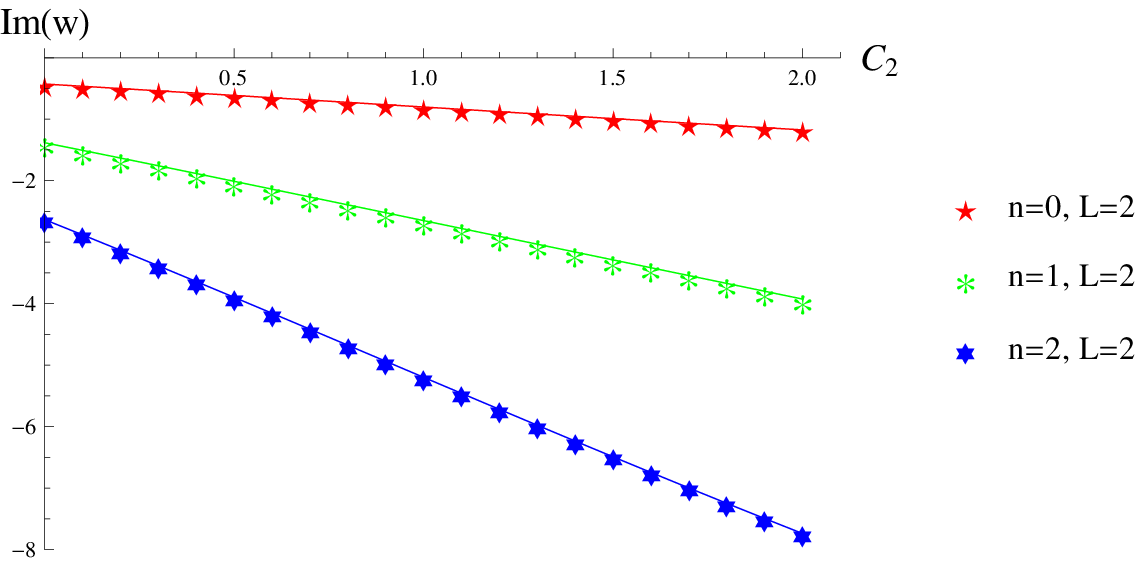} \par\medskip
\includegraphics[width=65 mm]{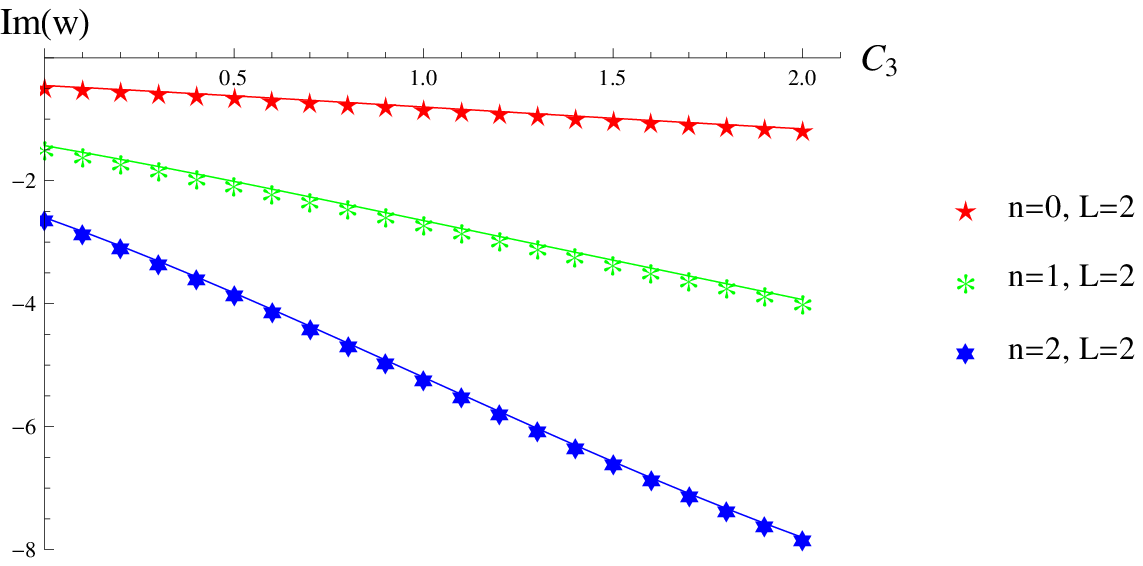} \hfil
\includegraphics[width=65 mm]{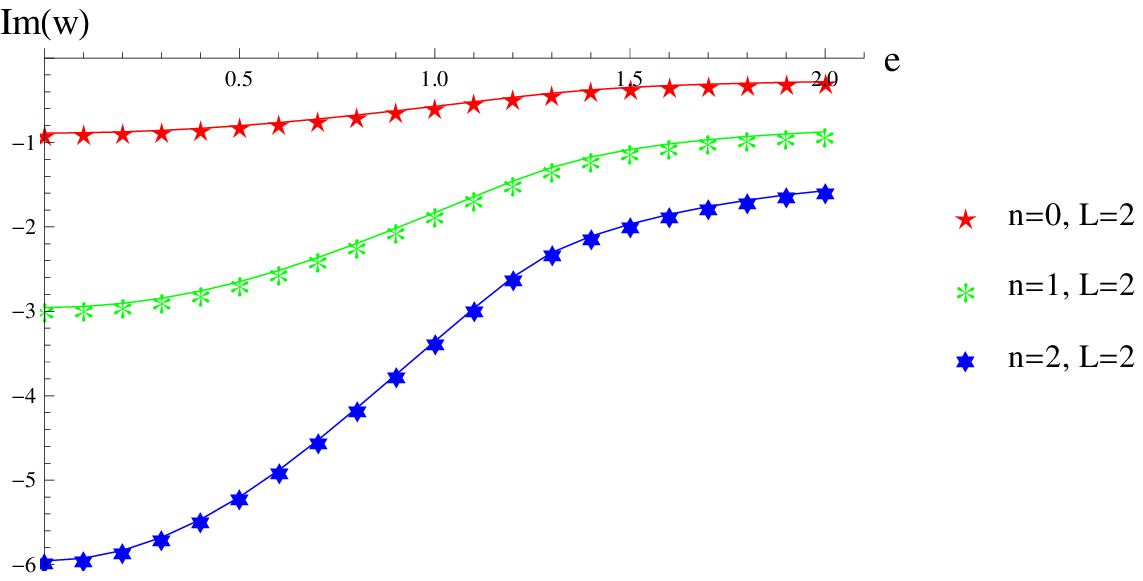}
 \caption{Imaginary part of scalar quasinormal modes with the same panel-coding as in Fig. \ref{real} }
    \label{imaginary}
\end{figure*}
\begin{figure*}[!htb]
    \centering
 \includegraphics[width=60 mm]{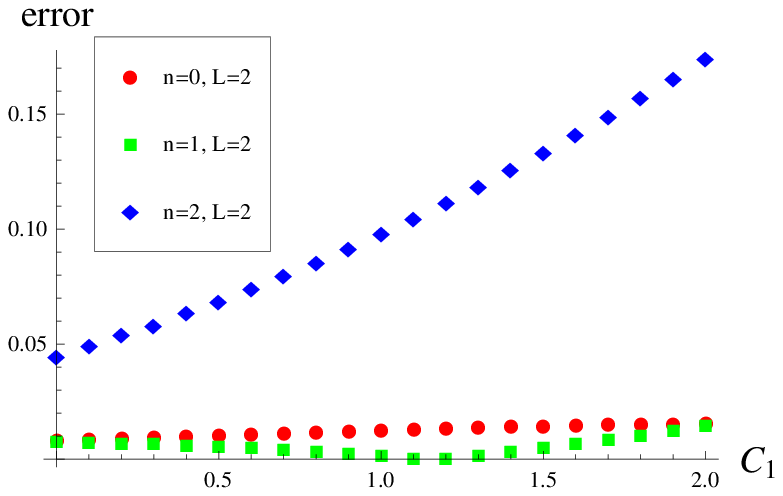} \hfil
\includegraphics[width=60 mm]{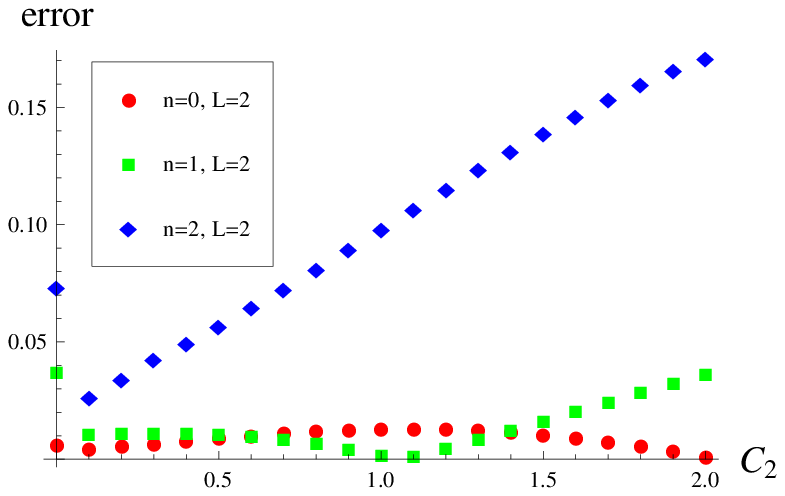} \par\medskip
\includegraphics[width=60 mm]{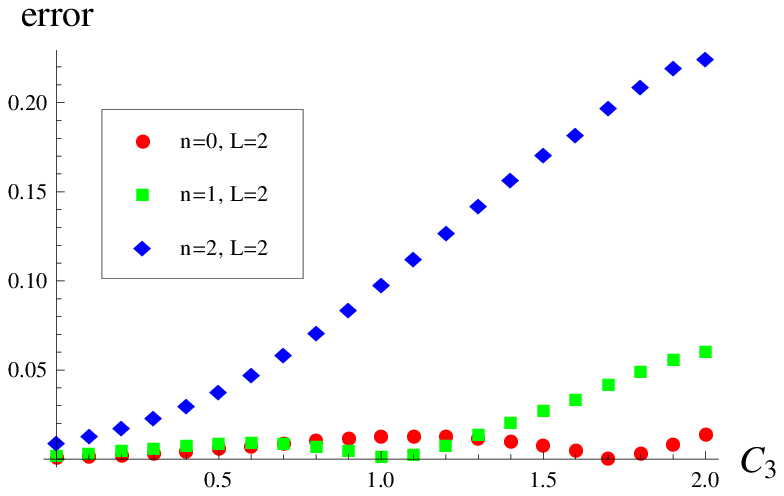} \hfil
\includegraphics[width=60 mm]{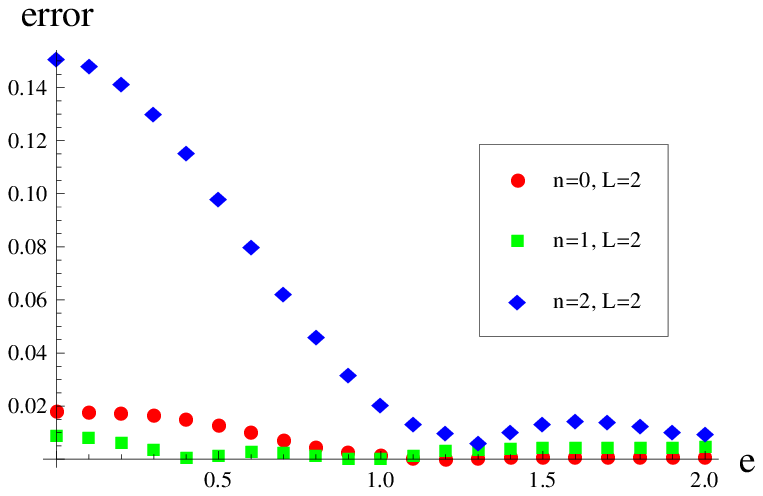}
 \caption{Error estimation of scalar quasinormal modes with the same panel-coding as in Fig. \ref{real}}
    \label{error}
\end{figure*}

In this section, we consider scalar perturbations around the black
hole to find the quasinormal mode frequencies. In this regard, we
use the WKB method which was first introduced by Schutz and Will
to the third order \cite{Schutz:1985km} and then extended to the
6th order \cite{Konoplya:2003ii} and recently to the 13th order
\cite{Konoplya:2019hlu}.

The equation of motion for a scalar field is given by $
\nabla^\mu\nabla_\mu \Phi=0 $. Using separation of variables, the
scalar field can be written as
$\Phi(x^\mu)=\sum\frac{\Psi_L(r)}{r}Y_{Lm}(\theta,\phi)e^{-i\omega
t}$ where $L$ denotes multipole quantum number and $ \omega $ is
generally a complex number with the real part as the frequency and
imaginary part as the decaying rate of QNMs. Regarding the
spherical symmetry of the metric, with the help of spherical
harmonics, one can write $\Delta
Y(\theta,\varphi)=-L(L+1)Y(\theta,\varphi)$.  Introducing the
tortoise coordinate $r_*=\int\frac{dr}{f(r)},$ it is possible to
write the radial part of the wave equation as follows
\begin{equation}
\label{QNMsEqu}
\frac{d^2\Psi}{dr_*^2}+[\omega^2-V(r)]\Psi(r_*)=0,
\end{equation}
where the effective potential is given as
\begin{equation}
V(r)=f(r)\Big( \frac{f'(r)}{r}+\frac{L(L+1)}{r^2} \Big).
\end{equation}

The WKB method is used for effective potentials with the form of a
potential barrier and constant values at the event horizon and
spatial infinity. Thus, we consider asymptotically flat metric to
have constant behavior at spatial infinity. The effective
potential for some values of parameters is plotted in Fig.
\ref{potential}. We see that the conditions are satisfied except
for the case of $ e=2 $ in the last panel with a potential
pit before the barrier. Although it may pose inaccuracy,
however, we also do calculation for this excepting case and we
back this subtlety after the error estimations found.

It should be noted that increasing the WKB order does not always
lead to a better approximation for the frequency, so we consider
calculations at the 6th order approximation. As an example, the
real and imaginary values of quasinormal modes for $ L=2 $ and $
n=0, 1, 2 $ are shown in Figs. \ref{real} and \ref{imaginary},
respectively. In both figures, each parameter is varied from $ 0-2
$ at fixed other parameters. We observe that as the massive
parameters $ \mathtt{C}_i ( i=1,2,3 )$ increase, the real and
imaginary (in magnitude) parts of the quasinormal frequencies grow
mostly on a straight line. It means that higher energy modes decay
faster. As an exception, the behavior of the real part for mode $
L=2, n=2 $ is not trivial as $ \mathtt{C}_3 $ increases and  Re
($\omega$)  is shrinking for variation of $ \mathtt{C}_3  $ from
$0-1$. It seems that as the effect of massive gravitons gets
stronger, the energy of the QNMs grows but, with a shorter
lifetime, they decay faster.

However, for constant massive parameters, we observe that Re
($\omega$)  and Im ($\omega$) (in magnitude) get lower with
increasing the Yang-Mills parameter while the real part of mode $
L=2, n=2 $ has growing behavior for $ e $ between $ 0 $ and $ 1.1
$. According to the figures, it seems that increasing ``$e$" leads
to appearing instability. In other words, one may find a critical
value of ``$e_c$" with respect to other parameters, in which the
solutions are stable for $e<e_c$.

We find that as the overtone number $ n $ increases, the scalar
perturbations have less energy for oscillation and decay faster
(with lower real and higher imaginary frequencies). From negative
values for the imaginary frequency, we conclude that we have
decaying oscillations, but as it is mentioned in Ref.
\cite{Konoplya:2019hlu}, it may not prove the stability of black
hole because the method of WKB approximation always leads to $ Im
(\omega) <0$.

In order to estimate the error of the WKB approximation, one can
compare two sequential orders. Here, we do this comparison between
orders of six and seven such that the error approximation is
defined by
\begin{equation}
\Delta \omega= \big\vert \dfrac{\omega_{7}-\omega_{6}}{\omega_{6}}\big\vert.
\end{equation}

The error values are given in Fig. \ref{error}. We observe that
the error estimation of $ L=2$ and $n=2 $ for $ \mathtt{C}_i>1 $
and $ 0<e<0.5 $ becomes larger than 0.1 and then it seems that
studying other known methods for obtaining quasinormal frequencies
in this region could be helpful.  Otherwise, the small values of
error estimation guarantee enough accuracy for results. Moreover,
from the last panel, we find the small error for $ e=2 $, though
the potential, in this case, is not as the proper form.


\section{Conclusion}\label{secCon}

In this paper, we considered the Einstein-Massive-Yang-Mills
gravity in $5-$dimensions. We obtained the thermodynamic
quantities for the planar AdS solution and showed that the first
law of thermodynamics has been satisfied. By defining a pressure
proportional to the cosmological constant, we performed the
criticality analysis in the extended phase space thermodynamics.
From the given equation of state, analytical relations for the
critical quantities were computed. We found that there exists just
one specific critical point for the given parameters which leads
to critical behavior, physically. By studying the effects of
Yang-Mills and massive parameters on the critical quantities, two
tables were given which showed that increasing the
Yang-Mills parameter $``e"$ leads to increasing $r_c$ and decreasing $%
\mathcal{T}_c$ and $P_c$. In addition, by increasing the massive parameters $%
\mathtt{C_2}$ and $\mathtt{C_3}$ the opposite behavior has been
observed. Moreover, since the critical quantities are independent
of $\mathtt{C}_1$, its variation does not affect the critical
behavior of the solution.

We also studied the phase transition and its properties through
some figures. According to the $P-r_{+}$ diagram, we found a first
order van der Waals like phase transition for
$\mathcal{T}<\mathcal{T}_c$. Furthermore, the existence of two
divergence points in $C_{p}-r_{+}$ plot and also the swallow-tail
behavior in the $G-T$ diagram are the characteristics of the first
order phase transition. All of these remarks guided us to find the
fact that the planar AdS black hole can exhibit the van der Waals
like phase transition between the small and large black holes. As
we showed such behavior comes from the massive term of the
solutions, while it does not occur for the planar solutions of the
Einstein-AdS and Einstein-Yang-Mills-AdS models.

We should note that in order to have the van der Waals phase
transition in Einstein-AdS, the existence of electric charge is
necessary, in addition to the spherical topology of the horizon
(see Ref. \cite{Dolan,Kubiznak-Mann}). Although there is no
abelian charge in our solution, the existence of massive term in
the Lagrangian is sufficient to guarantee the existence of van der
Waals like phase transition.

We continued by investigating the photon sphere and the shadow
observed by a distant observer. The results showed that the radius
of the photon sphere decreases with increasing the massive
parameters (decreasing the Yang-Mills parameter). We presented the
black hole shadow for different Yang-Mills and massive parameters.
It is observed that the shadow size shrinks with decreasing $``e"
$ and increasing $ \mathtt{C}_i$, which is the same as the results
obtained for variation of the photon sphere radius.

Finally, we calculated the quasinormal modes of scalar
perturbation for  $ L=2 $ and $ n=0, 1, 2 $. We found that as the
massive parameters $ \mathtt{C}_i ( i=1,2,3 )$ increase, the real
and imaginary (in magnitude) parts of the quasinormal frequencies
grow mostly on a straight line. But the behavior of the real part
for mode $ L=2, n=2 $ is not trivial as $ \mathtt{C}_3  $
increases and  Re ($\omega$)  is shrinking for variation of $
\mathtt{C}_3 $ from $0-1$. For constant massive parameters, we
observed that Re ($\omega$)  and Im ($\omega$) (in magnitude) get
lower with increasing the Yang-Mills parameter while the real part
of mode $ L=2, n=2 $ has growing behavior for $ e $ between $ 0 $
and $ 1.1 $.

As future work, it is interesting to find other black string/brane
solutions of the alternative theories of gravity with non-trivial
horizon topologies and discuss the possibility of the criticality
and phase transition. In addition, it will be useful to generalize
the obtained solution to higher dimensions and study the effect of
dimensionality.

\begin{acknowledgements}
We thank Shiraz University Research Council. This work was
supported in part by the Iran Science Elites Federation.
\end{acknowledgements}

\end{document}